**Non-invasive biomarkers of fetal brain development reflecting prenatal stress: an integrative multi-scale multi-species perspective on data collection and analysis**

Martin G. Frasch[1], Silvia Lobmaier[2], Tamara Stampalija[3], Paula Desplats[4], María Eugenia Pallarés[5], Verónica Pastor[5], Marcela Brocco[6], Hau-tieng Wu[7,8], Jay Schulkin[1], Christophe Herry[9], Andrew Seely[9], Gerlinde A.S. Metz[10], Yoram Louzoun[11], Marta Antonelli[5]

[1] Dept. of Obstetrics and Gynecology, University of Washington, Seattle, USA
[2] Frauenklinik und Poliklinik, Klinikum rechts der Isar, Technical University of Munich, Munich, Germany
[3] Unit of Fetal Medicine and Prenatal Diagnosis, Institute for Mother and Child Health IRCCS Burlo Garofolo, Trieste, Italy
[4] University of California, San Diego, Departments of Neurosciences and Pathology
[5] Instituto de Biología Celular y Neurociencia "Prof. Eduardo De Robertis", Facultad de Medicina, Universidad de Buenos Aires, Argentina
[6] Instituto de Investigaciones Biotecnológicas - Instituto Tecnológico de Chascomús (IIB-INTECH), Universidad Nacional de San Martín - Consejo Nacional de Investigaciones Científicas y Técnicas (UNSAM-CONICET), San Martín, Buenos Aires, Argentina
[7] Dept. of Mathematics and Dept. of Statistical Science, Duke University, Durham, NC, USA
[8] Dept. of Mathematics, University of Toronto, Toronto, ON, Canada
[9] Ottawa Hospital Research Institute, Ottawa, ON, Canada
[10] Canadian Centre for Behavioural Neuroscience, Department of Neuroscience, University of Lethbridge, Lethbridge, Alberta, Canada
[11] Bar-Ilan University, Dept. of Applied Mathematics, Israel

**Short title:** Biomarkers of fetal brain programming

**Corresponding author**
Martin G. Frasch
Department of Obstetrics and Gynecology
University of Washington
1959 NE Pacific St
Box 356460
Seattle, WA 98195
Phone: +1-206-543-5892
Fax: +1-206-543-3915
Email: mfrasch@uw.edu

**Abstract**
Prenatal stress (PS) impacts early postnatal behavioural and cognitive development. This process of 'fetal programming' is mediated by the effects of the prenatal experience on the developing hypothalamic-pituitary-adrenal (HPA) axis and autonomic nervous system (ANS). The HPA axis is a dynamic system regulating homeostasis, especially the stress response, and is highly sensitive to adverse early life experiences. We review the evidence for the effects of PS on fetal programming of the HPA axis and the ANS. We derive a multi-scale multi-species approach to devising preclinical and clinical studies to identify early non-invasively available pre- and postnatal biomarkers of these programming effects. Such approach would identify adverse postnatal brain developmental trajectories, a prerequisite for designing therapeutic interventions. The multiple scales include the biomarkers reflecting changes in the brain epigenome, metabolome, microbiome and the ANS activity gauged via an array of advanced non-invasively obtainable properties of fetal heart rate fluctuations. The proposed framework has the potential to reveal mechanistic links between maternal stress during pregnancy and changes across these physiological scales. Such biomarkers may hence be useful as early and non-invasive predictors of neurodevelopmental trajectories influenced by the PS. We conclude that studies into PS effects must be conducted on multiple scales derived from concerted observations in multiple animal models and human cohorts performed in an interactive and iterative manner and deploying machine learning for data synthesis, identification and validation of the best non-invasive biomarkers.

# INTRODUCTION

By 2010, 250 million children (43%) younger than 5 years in low-income and middle-income countries are at risk of not reaching their developmental potential.(Lu *et al.*, 2016; Black *et al.*, 2017) This is mostly due to exposure to biological and psychosocial factors that might alter brain function.(Grantham-McGregor *et al.*, 2007; Walker *et al.*, 2007) It is now widely accepted that maternal distress including depression, anxiety, stress, fears and worries have been identified as key risk factors affecting child development that requires urgent intervention.(Walker *et al.*, 2007) (Fontein-Kuipers *et al.*, 2014; Rakers *et al.*, 2017) Early postnatal care was shown to partially reverse the effects of prenatal stress (PS) on brain reprogramming in animal models.(Barros *et al.*, 2004; Weaver *et al.*, 2005; Barros *et al.*, 2006b) Early interventions promoting mother-infant bonding and cognitive stimulation, mainly based on education and maternal support programs and which are accessible in developing countries, might improve developmental outcomes in PS-exposed children, decreasing liability for psychopathology. (Fontein-Kuipers *et al.*, 2014; Nolvi *et al.*, 2016)

PS impacts early behavioral and cognitive development in human infants.(Mulder *et al.*, 2002; Beydoun & Saftlas, 2008) As a result, infants may develop attention-deficit hyperactivity disorder (ADHD) and sleep disturbances.(Weinstock, 1997) Longer-term persistence of these disorders may lead to depression and vulnerability to psychotic disorders in adulthood.(Cohen *et al.*, 1983; van Os & Selten, 1998) The underlying mechanisms of this fetal programming of adult diseases are thought to be mediated by the impact of PS on the developing hypothalamic-pituitary-adrenal (HPA) axis, an essential homeokinetic system capable of responding to stressors.(Van den Hove *et al.*, 2006) HPA axis is highly sensitive to adverse early life experiences(Meaney, 2001). Animal studies show that PS results in vulnerability to anxiety and impaired learning, memory and locomotor dysfunction.(Weinstock, 2001; Huizink *et al.*, 2004) Exposure to PS results in increased responsiveness of the HPA axis to stress, and reductions of glucocorticoid receptor (GR) expression in the hippocampus of adult offspring.(Zuena *et al.*, 2008) In humans, prenatal depressed/stressed maternal mood is associated with higher rates of preterm delivery and lower birth weight,(Wadhwa *et al.*, 1993; Van den Bergh *et al.*, 2005) elevated cortisol,(Field *et al.*, 2004) impaired subsequent working memory performance in young women (Entringer *et al.*, 2009) and changes in the epigenetic regulation of GR expression.(Mulder *et al.*, 1997)

A possible therapeutic avenue to counteract the effects of PS-induced fetal programming is given by postnatal stimulation. This can be accomplished by changes in the postnatal environment, such as care and early adoption. The positive effects of such treatment include better cognitive performance of adult offspring (Meaney *et al.*, 1988) and reduction of stress-induced corticosterone secretion.(Wallen *et al.*, 1999)

The first step in devising and testing early interventions to prevent PS effects on offspring is the identification of changes of the intrauterine environment using reliable and robust biomarkers of stress-related epigenetic reprogramming. We propose a conceptual multi-species multi-scale framework to discover early biomarkers of PS in the exposed infants. First, we review the experimental animal models and human cohort approaches for study of PS. We deliberately focus on two very different animal models: pregnant rat and sheep. The former lends itself to efficient studies of generational effects PS exerts on the offspring. The latter is *de facto* the only model of fetal physiology that allows fetal instrumentation and chronic monitoring with direct translational relevance to obstetrical practice. We conclude from this section of the review that a combination of such animal models and human cohorts into one systematic multi-species approach holds the key for a comprehensive and clinically relevant modeling of the PS effects. Second, we derive from these observations that such multi-species approach also needs to consider within one paradigm the multiple physiological scales of complexity which all are affected by the PS. Lastly, we review the mathematical instruments which are needed to tackle the complexity of data sets such studies would generate.

## *PS REWIRES THE BRAIN: A MULTI-SPECIES APPROACH*

Various mammalian species have been used to document the multifarious effects of PS on brain development and function ranging form rats to human cohorts. In spite of the wealth of careful retrospective and prospective studies on PS, there are still several confounding factors that cannot be controlled in human studies such as genetic and environmental factors, as well as the social environment.(Weinstock, 2001, 2008) For these reasons, studies in this field continue to rely on animal experiments, due mainly to their shorter life span and short breeding cycles, and because they offer the possibility to control the type, intensity, duration and timing of the stressor applied to the dam, the long term behavioral outcomes, as well as the interaction of the mother with her offspring in a controlled environment. The effects of PS on brain development in animal models have been mainly conducted in Sprague-Dawley, Wistar and Long-Evans rat strains, but also in Rhesus macaques, guinea pigs, sheep and mice.(Arnsten, 2000; Kapoor *et al.*, 2008; Weinstock, 2008; Braun *et al.*, 2017)

Most of the prenatal stress rodent models have been performed in rats and these studies have generated a large body of evidence towards the understanding of the mechanisms of developmental programming, especially in relation to the possibility of exploring the brain regions involved in neurogenesis and neuronal plasticity. (Fatima *et al.*, 2017)

Chronically instrumented non-anesthetized fetal sheep is an appropriate and uniquely suited animal model for studying the effects of in utero insults on fetal development, because of its recognized physiological and pathophysiological similarities to human fetal developmental profile and the unique ability to chronically instrument and monitor the fetus while manipulating the intrauterine environment.

Despite the diversity of findings, studies performed in human cohorts have shown the profound impact of PS on the cognitive development of the infants. However, research in humans has mainly been restricted to behavioural studies and to the macroscopic neuroanatomical analyses, while the genetic/epigenetic analysis has been limited to peripheral tissues due to the obvious inaccessibility of the brain. (Braun *et al.*, 2017)

The following will focus on the recent work in rodents, sheep and humans to provide an insight into some of the conceptual framework of mechanisms of perinatal and transgenerational programming.

*1) Studies in pregnant rat models of PS*

Up to date, the most comprehensive behavioural, morphological and histological information comes from studies in rodent models.(Boersma & Tamashiro, 2015; Weinstock, 2017) For many years, the outcomes were analysed in the first generation offspring only, but more recently a multigenerational paradigm has been established.(Babenko *et al.*, 2015)

*a) Single-generational studies*

In rodents, various PS protocols have been deployed ranging from saline injections, suspension, crowding, hypoxia, electric foot shock and placental insufficiency to unpredictable stress, noise and REM sleep deprivation.(Huizink *et al.*, 2004; Mastorci *et al.*, 2009) A frequently used protocol is a modified version of Ward and Weisz model consisting of restraining the mothers during the last week of gestation.(Ward & Weisz, 1984) This model induces a robust psychoneuroendocrine stress activation in the mothers.(Mastorci *et al.*, 2009) Abundant evidence demonstrates that exposure to different stressful events during the last week of pregnancy in rats interferes with the physiological progeny development inducing anomalies in neurogenesis and brain morphology that directly affect offspring behavior.(Weinstock, 2001; Mastorci *et al.*, 2009; Charil *et al.*, 2010) PS induces low birth weight, learning and attention deficits, impaired adaptation to stressful conditions, vulnerability to anxiety and depressive-like behaviors, reduced social interaction and some of the characteristic neuronal changes of schizophrenia.(Alonso *et al.*, 1991; Weinstock, 2001; Huizink *et al.*, 2004; Koenig *et al.*, 2005; Yang *et al.*, 2006; Yaka *et al.*, 2007; Darnaudery & Maccari, 2008; Weinstock, 2008) At the morphological level, there is a reduced dendritic arborization and astroglial hypertrophy with synaptic loss suggesting a possible alteration of glutamate metabolism. Glutamate transporters are altered in frontal cortex and hippocampus of PS offspring.(Barros *et al.*, 2006a; Adrover *et al.*, 2015)

Furthermore, enhanced propensity to self-administer drugs such as amphetamine and nicotine was observed in PS rats.(reviewed in (Pastor *et al.*, 2017)) PS induced delays in motor development and alterations in locomotor and exploratory activities that depend on the age and the sex of the offspring.(Henry *et al.*, 1995; Diaz *et al.*, 1997) PS also affects sexual behavior of adult offspring and gonadal dysfunction. Studies report a reduction in the number of adult male copulations, decreased number of ejaculations, enhanced lordotic-like behaviors and increased male partner preference over receptive females.(Shono & Suita, 2003; Gerardin *et al.*, 2005; Kapoor & Matthews, 2011) Additionally, a persistent lack of tonic gonadotropin secretion and altered testosterone secretion profiles were shown.(Shono & Suita, 2003; Gerardin *et al.*, 2005; Rodriguez *et al.*, 2007) Alterations in sexual morphological parameters, such as the anogenital distance length and the timing of testicular descent, were also found.(Barros *et al.*, 2004) Moreover, PS induces long-term imbalance of male sexual hormones concentrations in serum, advanced spermatogenesis development and age-dependent misbalance of alpha-receptor expression in prefrontal cortex and hippocampus.(Pallares *et al.*, 2013a; Pallares *et al.*, 2013b)

It is interesting to mention that most of the alterations described in PS animals are related to changes in midbrain dopaminergic activity, suggesting that the development of the dopamine (DA) is sensitive to disruption by exposure to early stressors. In fact, PS alters the asymmetry in D2 type receptors in the Nucleus Accumbens, an area associated with impulsivity (Fig 1).(Berger *et al.*, 2002; Barros *et al.*, 2004; Weaver *et al.*, 2005; Barros *et al.*, 2006b; Adrover *et al.*, 2007; Silvagni *et al.*, 2008; Carboni *et al.*, 2010; Katunar *et al.*, 2010) Adoption at birth reverses the receptor increase, reflecting the high vulnerability of DA system to variations both in prenatal and in postnatal environment.

Many studies have associated prenatal and perinatal early life adversity with changes in the global methylation status as well as at particular loci. For example, differential rearing (maternal versus surrogate-peer rearing) of rhesus macaques results in genome-wide differential methylation in both PFC and T cells.(Provencal *et al.*, 2012) Maternal behavior modified the methylation pattern in a region of the glucocorticoid receptor Nr3c1 promoter responsible for controlling its hippocampal expression.(Weaver *et al.*, 2004) Similarly, in hypothalamic neurons, maternal separation was associated with arginine vasopressin gene hypomethylation,(Murgatroyd *et al.*, 2009) and a decreased DNA methylation of the CRF gene was related to gestational stress.(Mueller & Bale, 2008) The expression of the gene *gpm6a* (which encodes membrane glycoprotein M6A participating in the establishment of neuronal morphology) is increased in the hippocampus of the PS rats' brains. The pattern of methylation in a CpG island of the first intron of the gene gpm6a was altered in rats subjected to PS.(Monteleone *et al.,* 2014) In mice, sex-specific changes were found in the levels of DNA methyltransferase (DNMT) 1 protein, the levels of histone H3 acetylation (AcH3) in the hippocampus, and serum corticosterone concentration. Prenatal stress also induces spatial memory deficits and epigenetic changes in the hippocampus associated with gene repression and heterochromatin conformation.(Benoit *et al.*, 2015)

b) *Multi-generational studies*

The lifetime risk of mental illness is greater if an individual has family history of a similar condition. Therefore, most efforts during the past decades have focused on the identification of genetic associations, but the results were often met with disappointment, as the associations with genetic variations were weak. In the post-genomic era, the focus has shifted to identify associations between epigenetic regulators that can be modified by environmental factors, such as stress. Importantly, epigenetic marks are potentially heritable. Trans-generational studies are uniquely suited to reveal mechanisms of trans-generational programming by inheritance of epigenetic, metabolomic and phenotypic traits. The multi-generational stress resembles human populations living in chronic stress conditions, *e.g.*, several generations exposed to residential school, war or poverty. (Laplante *et al.,* 2016) (Santavirta *et al.*, 2017)

Compared to single-generation PS, the multigenerational stress has somewhat different consequences: it facilitates adaptation to the recurrent maternal stress across generations, thus generating stress resilience. This causes new behavioral traits and better brain activity coherence. Based on the mismatch hypothesis, the *multi*generational stress leads to better adaptation because the offspring is bred for a stressful environment and the stress is indeed occurring again when the daughters get pregnant. The *trans*generational cohort faces the mismatch problem because they are bred for a stressful environment, but there is no more stress during pregnancy or any other time.

1. **PS elevates stress responses and the risk of mental illness**. PS (F1 generation) impedes developmental milestones in rats, thus desynchronizing brain development along with epigenetic signatures of human anxiety, depression, and adverse brain development.(Zucchi *et al.*, 2014) Interestingly, PS-induced anxiety-like and depression-like behaviours become most evident at the most vulnerable periods in life, early development and old age.(Erickson *et al.*, 2014)

2. **PS programs risk of anxiety and depression in future generations**. In a rat model of trans- and multigenerational experience, PS induces increased risk of gestational diabetes, preterm birth, and delayed brain development across generations. These manifestations are linked to microRNA (miRNA) and mRNA signatures of preterm birth(Yao *et al.*, 2014) and mental illness, in particular anxiety and depression-like symptoms, and altered brain connectivity in adulthood.(McCreary *et al.*, 2016) Interestingly, these studies revealed striking sex differences, with stressed females displaying partial stress resilience until the F3 generation, suggesting truly epigenetic inheritance.

3. **Stress-induced epigenetic and metabolic changes propagate across generations**. Stress alters miRNA expression patterns in the brain,(Babenko *et al.*, 2012) thus generating potentially heritable biomarkers of disease. New miRNA pathways have been identified which are involved in neurotrophin and myelin regulation, providing a mechanistic link to mental illness REF. Furthermore, altered epigenetic regulation of gene expression is also accompanied by altered metabolic footprints, which can be assessed in animals and humans using body hair, bio-fluids or solid tissues using NMR spectroscopy and inductively coupled plasma mass spectroscopy.(Ambeskovic *et al.*, 2013) These findings suggest that PS induces stable transgenerational specific epigenetic and metabolic alterations that can also be found in human disease.

MicroRNAs have become appreciated as a means for transgenerational epigenetic inheritance due to their small size, as they can translocate easily during meiosis and fertilization. By contrast, due to the epigenetic mark erasure in the germ line cells, the chromatin remodelling mechanisms represent a more controversial way of transmitting environmental cues across generations. However, there are some reports indicating that certain genomic regions are not demethylated. Thus, they could retain the information to be transmitted to descendants.(Lim & Brunet, 2013) Dias et al. have shown the transmission of odor aversion. They found in the sperm the hypomethylation in the Olfr151 gene that codes a known odor receptor.(Dias & Ressler, 2014) In addition, histone methylation at particular loci in the sperm can be affected by paternal diet and has been associated to an altered cholesterol and lipid metabolism in the offspring.(Carone *et al.*, 2010)

*2) Studies in pregnant sheep models of PS*

Fetal sheep and guinea pig (Iqbal *et al.*, 2012) in particular have been used extensively for studies of effects of antenatal synthetic glucocorticoid (sGC) treatment on fetal brain development. Many studies showed detrimental acute and transgenerational effects on neurodevelopment and HPA axis responsiveness to stress.(Schwab *et al.*, 2001; McCallum *et al.*, 2008; Antonow-Schlorke *et al.*, 2009; Iqbal *et al.*, 2012; Schwab *et al.*, 2012; Anegroaie *et al.*, 2016) It is not clear whether the postnatal brain can fully compensate for these changes. Consequently, despite the acute benefits of the antenatal sGC treatment to the fetus during labour and in the early postnatal period, further studies are needed to delineate the long-term effects of repeated sGC courses on postnatal brain development. For such fetal/postnatal experimental paradigms, the guinea pig model has been instrumental, although it is possible to do similar work in larger mammals such as sheep or non-human primate. To the extent that sGCs represent a stress stimulus to the fetus and can be injected directly intravenously to the fetus, this experimental approach also represents a possible paradigm for mimicking human fetal stress exposure to maternal stress hormones without accounting for the interindividual and interspecies differences in the placental transfer dynamics. A pharmacological form of PS, antenatal GCs alter the set point of the HPA axis, which matures during late gestation. Taken together, as an iatrogenic stressor or a model for stress-induced fetal programming, sGC-driven studies in guinea pig and sheep have shown the potential of fetal stress exposure to alter organ development, in particular that of the brain.(Moisiadis & Matthews, 2014b, a)

A more "human-like", but also technically more complex experimental paradigm involves isolation of pregnant ewes based on the fact that they are flock animals and experience such isolation as stress. Such approach results in acute and chronic stress-induced adaptations and represents the most comprehensive animal experimental model of human *fetal* stress exposure.(Rakers *et al.*, 2013)

Taken together, these results suggest that PS insults are critical in the development of biochemical responses and behavior in adults, and that maternal care is crucial both during pregnancy and in the first weeks of life. (Fontein-Kuipers *et al.*, 2014) (Nolvi *et al.*, 2016) It has been postulated that several psychiatric disorders that manifest themselves in the adult human, such as schizophrenia, depression, anxiety and drug abuse, are imbalances of dopaminergic, glutamatergic and GABAergic systems as a consequence, among other reasons, of alterations in the early development of the corticostriatal pathway. Rat models of gestational stress will provide clues to understanding the mechanisms by which a PS insult in early life contributes to the breakdown of the balance in neurotransmission and the formation of aberrant cortical connections, which would entail the establishment of abnormal cognitive behaviors.

*3) Studies in humans*

There is now a large consensus that different types of PS in pregnant women are associated with altered outcomes for the child. Types of stress include maternal anxiety and depression, bereavement, daily hassles, bad relationship with the partner, and exposure to acute man-made or natural disasters. Several independent retrospective and prospective studies (Glover, 2015; Silveira & Manfro, 2015) have shown that PS is associated with lower birthweight and reduced gestational age,(Wadhwa *et al.*, 2011) a poorer performance on the Neonatal Behavioral Assessment Scale,(Rieger *et al.*, 2004) more difficult temperament,(Davis *et al.*, 2007; Werner *et al.*, 2007) sleep problems,(O'Connor *et al.*, 2007) lower cognitive performance and increased fearfulness associated with higher maternal stress during pregnancy.(Bergman *et al.*, 2007)

A well-established cohort is based on the 1998 Quebec Ice Storm which follows the consequences of maternal stress induced by this natural disaster. By studying natural disasters, such as the Project Ice Storm, the impact of maternal objective and subjective distress on genetic and epigenetic biomarkers can be estimated.(Cao-Lei *et al.*, 2014) Project Ice Storm revealed correlation between exposure to prenatal stress and differential DNA methylation of 957 genes in 13-year old.(Cao-Lei *et al.*, 2016a; Cao-Lei *et al.*, 2016b) The majority of the differentially methylated genes were related to immune function and metabolism. The methylation patterns seemed to mainly correspond to the degree of objective maternal stress rather than subjective stress reported by the mothers during pregnancy.

There are other human cohorts that allowed studies across the generations. In the Överkalix population in northern Sweden, the researchers found a relation between grandparent food availability and the grandchild's longevity. A food excess at ages 9-12 years of grandfathers correlated with short survival of grandsons. These effects might be triggered by methylation of epigenetic marks.(Bygren *et al.*, 2001) Famines during the first and second war (Germany, 1916-18(Van den Berg & Pinger, 2014) and Amsterdam, 1944-45(Roseboom *et al.*, 2001) also showed that exposure to adverse environment during the early developmental stages changes the outcomes of the next generations.

In sum, there is large body of evidence for the complexity of PS effects on the individual's physiology and the heterogeneity in the stress responses. Because of the interplay between genes and environment, finding PS biomarkers requires a multispecies approach. Rodent models serve to obtain single and transgenerational markers, sheep fetuses resemble closer human physiology and allow *in utero* monitoring while the human cohorts allow to analyze the reliability of the putative biomarkers.

**A MULTI-SCALE APPROACH TO DISCOVERY OF BIOMARKERS AND TREATMENT STRATEGIES**

Epigenetic markers may be correlated with maternal stress, depression and anxiety and with infants' cognitive development thus serving as novel biomarkers of PS. The animal model-centered review of the PS effects points to the impact of the PS on multiple physiological scales, from molecular level to complex system's level patterns. This dictates that to derive meaningful, translational biomarkers from preclinical and clinical studies, PS effects should be studied on those molecular and integrative levels in a unified multi-scale paradigm. Such unification can be achieved when multiple pertinent animal models and human cohorts studied are designed in concert, rather than as separate studies. In following we review the physiological scales relevant to gauging the PS effects comprehensively. We propose that such approach will yield clinically relevant PS biomarkers. Aside of biomarkers discovery, studying all physiological scales combined holds the potential to provide insights into therapeutic interventions to recover the PS brain phenotype. Such multi-scale paradigm requires novel mathematical methods of pattern discovery and integration. As we conclude below, rapid developments in machine learning hold the key to this methodology.(Marschik *et al.*, 2017)

*1) PS influences brain development epigenetically*

Studies on single, multi and transgenerational stress inheritance mechanisms have been conducted mostly using pregnant rat model of inescapable stress.(Monteleone *et al.*, 2014; Yao *et al.*, 2014) Other studies (reviewed in (Ho & Burggren, 2010; Blaze & Roth, 2015)) used maternal separation (Pusalkar *et al.*, 2016) or alterations in maternal behavior (Weaver *et al.*, 2004) or diet (Berardino *et al.*, 2017) as early-life stressors and found changes in DNA methylation, histone modifications and microRNA expression.

We propose that PS may result in patterns of co-variation between DNA methylation and levels of microRNA between brain, blood and saliva in rodent and sheep models of PS. This would serve as a model to validate saliva as the clinically easily accessible peripheral fluid serving as a biomarker of PS exposure and to correlate methylation levels with behavioral outcomes and stress responsiveness.

*a)     PS inheritance via DNA methylation*

Epigenetic changes can persistently alter gene transcription affecting physiology and behavior and are thought to underlie these long-term effects of PS.(Weaver *et al.*, 2004; Caldji *et al.*, 2011; Mulligan *et al.*, 2012) Increased HPA stress reactivity in the offspring of low maternal care rats is associated with higher DNA methylation at the promoter of *NR3C1* (which encodes GR).(Liu *et al.*, 1997; Francis *et al.*, 1999; Weaver *et al.*, 2004) More recently, Braithwaite et al, (2015) reported that maternal prenatal depressive symptoms significantly predicted increased *NR3C1* 1F DNA methylation in buccal cells of male infants.(Braithwaite *et al.*, 2015) In mice, levels of both OGT (O-linked-N-acetylglucosamine (O-GlcNAc) transferase) and its biochemical mark, O-GlcNAcylation, were significantly lower in males and further reduced by prenatal stress.(Howerton *et al.*, 2013) In humans, differential methylation is associated with prenatal exposure to maternal depression,(O'Connor *et al.*, 2003; Teh *et al.*, 2014) PS and birth weight.(Filiberto *et al.*, 2011; Mulligan *et al.*, 2012; Vidal *et al.*, 2014)

Preconceptual or intra-gestational stress may result in increased cerebral and placental expressions of the corticotropin-releasing hormone (CRH) gene stimulating fetal cortisol and adrenocorticotropic hormone (ACTH) and signaling premature maturation of fetal tissue.(Horan *et al.*, 2000; Moog *et al.*, 2016) Repeated stress exposure may dysregulate HPA axis and increase CRH and cortisol levels which in turn sensitizes women to stress experienced during pregnancy. Pre-gestational stress increased the expression of corticotropin-releasing hormone type 1 (CRH1) messenger RNA in the brains of mothers and offspring, suggesting an epigenetic route of transgenerational transmission.(Zaidan *et al.*, 2013) Pre-gestational stress to female rats two weeks prior to mating resulted in reduced anxiety, enhanced fear learning, and improved adaptive learning for second generation offspring.(Zaidan & Gaisler-Salomon, 2015) Levels of the stress hormone corticosterone (an indicator of HPA axis functioning) were altered across the three generations in a sex-dependent manner.(Zaidan & Gaisler-Salomon, 2015) Maternal stress during the third, but not the second, week of gestation in rats was associated with alterations in stress reactivity behaviors and prolonged elevations in glucocorticoid levels among adult male offspring.(Koenig *et al.*, 2005) Heightened anxiety was associated with greater CRH mRNA gene expression in the amygdala, and attenuated stress responses were associated with greater glucocorticoid mRNA expression in the hippocampus and impaired feedback to the HPA axis.(Grundwald & Brunton, 2015) The offspring of rats exposed to either a daily injection of corticosterone or prenatal stress during the third week of gestation all displayed decreased GR protein levels in the medial prefrontal cortex, hippocampus, and hypothalamus, as compared to controls.(Bingham *et al.*, 2013) Similarly to the pre-gestational dysregulatory effects of stress on responsiveness to stress during pregnancy, increased CRH levels during stressful pregnancy act on the CRH receptor 1 (CRH-R1) to mediate increased maternal vulnerability after delivery with a suppressed HPA axis increasing the risk for postpartum depression.(Meltzer-Brody *et al.*, 2011; Engineer *et al.*, 2013) Overall, changes in maternal and offspring HPA axis function are modified via stress-induced changes to CRH expression (Zaidan *et al.*, 2013) and often accompanied by behavioral effects.

The studies focused on the HPA-axis have overlooked other physiological system affected by PS, such as the autonomic nervous system (ANS). In fact, (Bleker *et al.*, 2017) suggested that psychosocial stress in pregnancy might program the fetus through other mechanisms than through altering maternal cortisol levels. Moreover, (Braithwaite *et al.*, 2015) found no association between maternal cortisol and infant DNA methylation suggesting that the effects of maternal depression may not be mediated directly by glucocorticoids; instead, sympathetic nervous system activity, a component of the fetal ANS, may be the mediating pathway.

The following genes have been implicated in the stress response in relation to the HPA axis (paragraphs A,B,C,D) and to the ANS (E,F,G).

A) exon $1_7$ of *nr3c1* encodes the glucocorticoid receptor (GR), which can function both as a transcription factor that binds to glucocorticoid response elements (GRE sites) in the promoters of glucocorticoid responsive genes to activate their transcription and as a regulator of other transcription factors. It has been associated to a lower GR expression in hippocampus and with an exacerbated response to stress.(Weaver *et al.*, 2004; Murgatroyd *et al.*, 2009; Hackman *et al.*, 2010; Kertes *et al.*, 2016)

B) Intron 1 of *FKBP5* that contains GRE sites. The protein FKBP51 belongs to the immunophilin protein family, which plays a role in immunoregulation and basic cellular processes involving protein folding and trafficking. FKBP51 functions as a co-chaperone that interacts with the GR protein. Chronically-administered glucocorticoids reduce methylation at *FKBP5* locus.(Lee *et al.*, 2010) In humans, war trauma modified methylation of this gene.(Kertes *et al.*, 2016)

C) *HSD11B2* encodes the enzyme corticosteroid 11-beta-dehydrogenase, a microsomal enzyme complex responsible for the interconversion of cortisol and cortisone.

D) *CRH* encodes the corticotrophin releasing hormone (CRH), one of the major HPA regulators. Changes in *CRH* methylation have been associated with chronic stress in animal models(Mueller & Bale, 2008) and in humans.(Kertes *et al.*, 2016)

E) *GNAS1* encodes the alpha-subunit of the G-protein. This GNAS1 subunit has been associated with ANS(Yasuda *et al.*, 2004) and, more recently, with prenatal maternal stress.(Vangeel *et al.*, 2015).

F) *ELP1/IKBKAP* encodes a scaffold protein and a regulator for three different kinases involved in proinflammatory signaling. This protein can bind NF-kappa-B-inducing kinase (NIK) and IKKs through separate domains and assemble them into an active kinase complex. Mutations in this gene have been associated with familial dysautonomia.(Jackson *et al.*, 2014)

G) *IGF2*: this a member of the insulin family of growth factors, involved in development and growth. An association was reported between DNA methylation in one *IGF2* DRM and pregnancy-related anxiety.(Vangeel *et al.*, 2015)

    *b)*     *PS inheritance via miRNA signaling*

Experimental paradigms of PS in animal or human cohorts allow for blood sampling and extraction of T cells and PBMCs to analyze microRNA profiles from cellular RNA. These microRNA patterns can be compared to those measured in animal studies yielding data from the relevant brain structures such as prefrontal cortex and dentate gyrus of the hippocampus. PS modified the expression of several microRNAs in the hippocampus and prefrontal cortex of prepubertal and adult rat offspring, microRNA-133b being altered most significantly.(Monteleone *et al.*, 2014)

In animal studies, it is important to account for microRNA spatial expression variation and co-localization. microRNA targets of interest are those related to HPA axis function such as GR, MR, 11ß-HSD type 1 and type 2, FKBP5, STAT5B and MHC II, Hsp70 and Hsp90, and to neuronal plasticity and psychopathologies, such as cortical BDNF and glial cell-derived neurotrophic factor (GDNF). Previous analyses have shown that differentially regulated microRNAs included miR-34 (anxiety), and miR-132, 142-5p, 146a, 181b, 486-5p, 650 (depression), miR-124 (regulates GR expression)(Babenko *et al.*, 2015) and the miR-200 family.

## *2) PS and metabolome*

Even mild maternal stress induces epigenetic and metabolomic alterations across four subsequent generations of rats. Notably, many of the epigenetic and metabolic signatures altered by transgenerational stress in this rat model have been also identified as markers of mental illness in humans.(Zucchi *et al.*, 2014)

Future studies can expand these findings with deep sequencing and $^1$H nuclear magnetic resonance (NMR) spectroscopy to identify DNA methylation and microRNA signatures linked to impaired mental health using blood across species such as human, fetal sheep and rat cohorts and *link epigenetic and metabolomic profiles to endocrine markers of elevated stress response and adverse mental health outcomes*. The multi-species approach would allow to search for similarities between metabolomic patterns to identify possibly predictive epigenetic and metabolomic signatures of PS and transgenerational inheritance that are phylogenetically preserved. Corresponding epigenetic, genetic, behavioral and pathophysiological data can then be correlated with metabolomic outcomes. Determining metabolic linkages to brain development, mental health and wellness outcomes throughout the life-span and across generations has the potential to revolutionize the future of health care by transforming the current trends of curative care to personalized and preventive medicine.

We speculate that prenatal and transgenerational stress, through altered epigenetic regulation, programs the maternal, infant and child stress response and lifetime mental health trajectories. We predict that stress response and mental health status will be associated with distinct metabolic signatures in clinically accessible tissues such as saliva or placenta.

*3) PS and microbiome*

A surprising recent result that may help understand PS and perhaps even allow monitoring it involved the gut microbiome. The discovery of the placental microbiome re-fueled the debate whether fetus is exposed to and interacts with bacteria during development.(Aagaard *et al.*, 2014) It remains controversial whether the fetal compartment is colonized,(Boersma & Tamashiro, 2015) but the notion continues to attract attention, because such physiological mechanism would have profound impact on brain-gut communication via the vagus nerve(Liu *et al.*, 2015) hence influencing the fetal brain development.(Garzoni *et al.*, 2013; Leclercq *et al.*, 2017) Gut microbiota are essential to human health, playing a major role in the bidirectional communication between gut and brain.(Borre *et al.*, 2014; Haberman *et al.*, 2014) The significance and influence of the fetal intestinal microbiome on stress responses, epigenetic modifications and brain development remain to be explored. Interactions between the microbial community and the developing brain likely contribute to pathological brain development after birth.(Borre *et al.*, 2014; Haberman *et al.*, 2014) Future studies will test PS effects on microbiomes of fetal gut and placenta in animal studies and in human placenta (clinical cohorts) to derive predictive biomarkers of PS.

*4) PS and ANS*

Most PS-induced alterations have been described for hippocampal and prefrontal cortex neurons.(Fujioka *et al.*, 2006; Negron-Oyarzo *et al.*, 2015) However, changes in the morphology and the connectivity of the autonomic nervous system (ANS) neurons due to PS have been poorly studied. Like the cortical neurons, those from the ANS may also be affected during PS exposure. In the guinea pig, it has been observed that enteric neurons (i.e., peripheral ANS neurons) respond to CRH.(Liu *et al.*, 2005) Patients suffering from panic disorder provide a clinical model of stress. These patients show changes in the sympathetic nervous system also observed in patients with essential hypertension. A reduced neuronal noradrenaline reuptake is present in both disorders and epigenetic changes mediate them.(Esler *et al.*, 2008)

During pregnancy, two lines of investigations have indicated a role of the ANS in mediation of stress effects on fetal physiology and development.

First, maternal corticosteroid administration during pregnancy - frequently used for fetal lung maturation in cases of threatening preterm delivery and an iatrogenically administered pharmacological stressor - has shown to affect autonomic balance *in utero*.(Dawes *et al.*, 1994; Derks *et al.*, 1995; Mulder *et al.*, 1997; Senat *et al.*, 1998) This effect is transient, but repeated fetal administration of betamethasone alters nervous system maturation.

Second, the vagus nerve influences brain function and body metabolism in a pleiotropic manner.(Pavlov & Tracey, 2012, 2015) A new field of bioelectronic medicine is emerging. It aims to devise therapeutic approaches using vagus nerve stimulation (VNS) to modify the endogenous salutory signaling of the vagus nerve.(Borovikova *et al.*, 2000; Kwan *et al.*, 2016; Pavlov & Tracey, 2017) VNS reduced sympathetic tone, stress-induced anxiety behaviors and depression symptoms in animal models and in clinical studies.(O'Keane *et al.*, 2005; George *et al.*, 2008; Caliskan & Albrecht, 2013; Liu *et al.*, 2013; Clancy *et al.*, 2014; Pena *et al.*, 2014; Ylikoski *et al.*, 2017) VNS is thought to facilitate tonic inhibition of the basolateral amygdala by the infralimbic region of the medial prefrontal cortex, which results in reduced fear response.(Caliskan & Albrecht, 2013) VNS increases CRH expression in hypothalamus(Hosoi *et al.*, 2000) and CRH receptor 1 agonism increases vagal modulation of HRV.(Porges, 1995; Farrokhi *et al.*, 2007; Porges, 2009) This reciprocal CRH - vagus nerve circuitry provides an important diagnostic and therapeutic link between stress and the ANS.

Notably, novel non-invasive methods of VNS are being developed which will not require surgical cervical VNS implants, have minimal to no side effects, and are low-cost.(Liu *et al.*, 2013; Clancy *et al.*, 2014; Frangos *et al.*, 2015; Ylikoski *et al.*, 2017) It is now possible to conceive of VNS treatment of neonates.

Together, there is strong evidence that vagus nerve activity is a key player in PS-induced brain programming, can be monitored using innovative fetal heart rate (FHR) analysis techniques (reviewed in detail below) and used as endogenous homeostatic mechanism to potentially recover the PS induced phenotype early postnatally. This offers another pillar of interventions, complementary to the neurobehavioural strategies such as enrichment.

Fetal ANS function can be studied longitudinally using FHR analyses to measure biomarkers of PS-induced epigenetic reprogramming in human fetuses. Mothers identified as having been "stressed" and controls can be monitored with trans-abdominal non-invasive fetal and maternal ECG (ta-fECG; ta-mECG) for ANS assessment. Infants' cognitive development can be assessed by the Bayley Scale III of Infant development at 18 months of age. This approach also permits quantification and correlation of ANS and behavioral data to the epigenetic biomarkers from salivary DNA obtained from the neonates and young infants.

Advanced FHR monitoring techniques such as phase-rectified signal averaging (PRSA) or multidimensional FHR analysis are sensitive to detecting an impairment of fetal ANS. (Huhn *et al.*, 2011; Graatsma *et al.*, 2012; Lobmaier *et al.*, 2012; Casati *et al.*, 2014; Frasch *et al.*, 2014; Rivolta *et al.*, 2014; Li *et al.*, 2015; Stampalija *et al.*, 2015) Future studies will test their ability to identify fetuses affected by PS.

Advanced analysis of FHR patterns specifically assessing changes in the autonomic regulation of FHR may identify fetus at increased risk for pathological fetal programming. Early signs of hypoxemia are found in changes in the autonomic regulation of the FHR. This can be assessed by the relatively new PRSA method measured by cardiotocography (CTG) or electrocardiography (ECG),(Bauer *et al.*, 2006; Kantelhardt *et al.*, 2007) even in fetuses.(Stampalija *et al.*, 2016)

*a)    Phase-rectified signal averaging method*

Initially, PRSA has been described in adult cardiology for prediction of mortality after myocardial infarction and has been found to be superior to other methods.(Bauer *et al.*, 2006) PRSA can eliminate signal artifacts and noise and extract areas of interest. In contrast to other methods of analysis of FHR variability, PRSA permits the detection of quasi-periodicities in non-stationary data. PRSA has been successfully applied in fetal medicine, despite the challenges of a non-stationary signal, with more disturbance in the signal than in the adult after a myocardial infarction. The novel parameter referred to as cardiac average acceleration and deceleration capacity is more specific than the conventional FHR analyses (e.g. computerized cardiotocography and short term variation) in identifying intrauterine growth restriction (IUGR)

antepartum(Huhn *et al.*, 2011; Graatsma *et al.*, 2012; Lobmaier *et al.*, 2012; Lobmaier *et al.*, 2016) and strongly correlates with acid-base biomarkers during acute hypoxic stress in humans during labour(Georgieva *et al.*, 2014) and the fetal sheep model.(Rivolta *et al.*, 2014) Even more interestingly, it has been shown that IUGR fetuses with brain sparing (fetal adaptive mechanism to chronic hypoxemia) have a lower acceleration and deceleration capacities than growth restricted foetuses without brain sparing.(Stampalija *et al.*, 2016) This intimate inter-relation between brain perfusion and FHR is thought to be mediated via ANS (aortic chemoreceptors and carotid baro- and chemoreceptors). Newer data also show an activation of ANS in fetuses affected by maternal gestational diabetes which could not be seen using conventional techniques.(Lobmaier *et al.*, (2017))

To evaluate the ANS influence on FHR the beat-to-beat information (R-R intervals) should be analysed. The new generation of the trans-abdominal fetal ECG monitors (such as Monica AN24, Monica Healthcare, Nottingham, UK) allow for a completely non-invasive and passive recording of fetal and maternal ECG: it only records electrophysiological signals from the women's abdomen without hampering mobility or other diagnostic procedures.(Stampalija *et al.*, 2012) This signal can be then used for a more sophisticated analysis of FHR such as PRSA or the multidimensional FHR analysis.

b) *Maternal-fetal heart rate entrainment and multidimensional FHR variability analysis in fetal sheep and human cohorts*

Although the evidence of maternal-fetal heart rate entrainment, also referred to as synchronization, has been demonstrated,(Van Leeuwen *et al.*, 2009) its clinical potential as easily obtainable diagnostic or prognostic tool has remained untapped. We propose that both approaches should be explored both in large animal models and clinical studies to test their potential to predict maternal and fetal stress.

Complex signals bioinformatics approaches have been recently developed(Herry *et al.*, 2016) that will allow examination of the putative correlations between the measures derived from all heart rate analyses techniques and epigenetic markers, based on the assumption that PS imprints both phenotypic modalities permitting a mutual inference.

HRV analysis can be performed via a series of automated algorithms that process a waveform recording into a comprehensive multivariate characterization of its degree of variability and complexity (See for example, the Continuous Individualized Multiorgan Variability Analysis (CIMVA) software engine, Fig. 3).(Seely & Newman, 2016) First, individual heartbeats are identified from the ECG waveform, using commonly used QRS delineation algorithms a time series of R-peak to R-peak time intervals (RRI) is formed. A thorough automated assessment is performed on the quality of the ECG signal and RRI time series. Movement artefacts, noise, disconnections and saturations are identified on the ECG waveform. A beat-by-beat signal quality index can be derived, using continuity and morphology analyses. In addition, the RRI time series is filtered to exclude or correct non-sinus beats and non-physiologically plausible data. The signal complexity and degree of variability are then assessed using the cleaned RRI time series.

FHR variability monitoring requires tracking HRV over time and a moving window analysis is typically employed, whereby a window of fixed duration (or fixed number of RR intervals) is shifted in time across the entire duration of the RRI time series. A comprehensive set of linear and nonlinear variability metrics are calculated within each window, as each technique provides a unique perspective on the data and no single method can provide a complete characterization of the biologic signals. Rather, a combination of multiple techniques stands to deliver the most complete evaluation (Table 1; for detailed description of the HRV measures see Table 2).(Goldberger *et al.*, 2002; Bravi *et al.*, 2011) Variability metrics include measures characterizing the statistical properties (e.g. standard deviation, RMSSD), the informational complexity (e.g. entropy measures), the pattern of variations across time scales (e.g. fractal measures, power law exponents) or the energy contained in the signal (e.g. spectral measures). Only high quality variability estimates are used in subsequent modelling. The output of the FHR variability analysis is a multivariate representation of variability tracked over time, where the temporal relation between subsets of fHRV measures can help characterize the fetal innate immune system's response to endotoxin and monitor fetal inflammatory response. For example, in a fetal sheep model of inflammation Durosier et al.(Durosier *et al.*, 2015) calculated a large set of fHRV measures to track variability changes and the impact of LPS injection and resulting inflammation over time. Using population-based Principal Component Analysis (PCA) derived from LPS-injected animals, animal-specific fetal HRV temporal profiles were created, which tracked pro-inflammatory cytokine IL-6 profiles (Fig. 2).

In the fetal sheep model, FHR variability reflects maturation and activation of the parasympathetic branch of the ANS involved in sensing and control of fetal acidemia, hypoxia and inflammation.(Frasch *et al.*, 2007; Frasch *et al.*, 2009; Durosier *et al.*, 2013) In human cohorts and in the fetal sheep model of human labor and fetal inflammation, multidimensional FHR variability analysis can predict clinical outcomes after birth.(Liu *et al.*, 1997; Durosier *et al.*, 2013) In elephants, a highly complex social species with brains similar to humans, HRV-based techniques have been suggested to distinguish stressed versus non-stressed animals.(Vezina-Audette *et al.*, 2016) To build such predictive FHR acquisition systems, certain types of FHR monitors are required, such as the AN24 monitor. Such monitors have the advantage over traditional ultrasound now used for FHR monitoring in that they sample FHR at a frequency high enough (900 Hz) to enable detection of the more subtle fluctuations of FHR variability which reflect integrative pathophysiological fetal responses such as those to acidemia and likely also to stress.(Durosier *et al.*, 2014; Frasch *et al.*, 2014)

**MACHINE LEARNING APPROACHES**

The advance of technology allows us more convenient ways to observe the world, and accumulate more data from the world. The relationship between PS and how it impacts early postnatal behavioural and cognitive development is not an exception. We could now easily collect data, which could be direct or indirect measurements of the PS, for example, the DNA methylation, metabolome, microbiome, and ANS activity. While collecting data is relatively easy, without a proper data analysis, we cannot gain any benefit from the data. Machine learning is a fancy name for statistical tools aiming at such a data analysis (Friedman *et al.*, 2001). Conversely, to properly apply these tools, understanding the data is inevitable. The available data for PS have at least the following properties: data are collected from multimodal equipment and are of heterogeneous types, the system under observation for the PS is nonlinear and nonstationary, the data volume could be large and inconsistent across different facilities. Moreover, data quality could vary greatly and data collection velocity could be unpredictable. Based on these facts, choosing proper machine learning tools that are accurate in the prediction, stable to noise and computationally affordable could be challenging. New tools might need to be developed. We summarize an overall machine learning framework in Figure 4.

There are two main steps in machine learning: feature extraction and learning/regression. While there are several measurements available from different aspects for the PS, designing and selecting proper features from these measurements (Guyon & Elisseeff, 2003) is the key step toward successful machine learning. A typical challenge is the combination of heterogeneous data types, such as time-series, imaging, microbiome gene expression profile (OTU frequencies), and blood biomarkers. New learning methods will be required to develop multi-modal learning algorithms. Based on the needs to capture the nonlinearity and nonstationarity, a manifold learning technique called alternating diffusion, based on the low-dimensional geometric structure assumption, has been shown to be useful in fusing information from different modalities in the nonlinear fashion. It allows us to preserve the nonlinear/nonstationary underlying structure and remove the sensor-specific unwanted nuisance, before the learning procedure is applied (see, for example (Talmon & Wu, (2017))). Another typical challenge is the dimension reduction: as the data gets complicated and measurement gets more diverse, we need a more sophisticated way to identify those useful parameters/features and guarantee numerical. In addition to the traditional linear approaches as the principal component analysis, several nonlinear tools have been developed for this purpose, such as the locally linear embedding (Roweis & Saul, 2000), ISOMAP (Tenenbaum *et al.*, 2000), diffusion map (Coifman *et al.*, 2005), and so on. Sometimes this kind of dimensional reduction algorithm is called unsupervised learning. There are many learning techniques available, ranging from the traditional linear/logistic regression to the modern support vector machine, boosting and random forest, etc. (Friedman *et al.*, 2001) We call these learning techniques the supervised learning, which means that we are learning the system based on the experts' labeling or truth. How to choose a proper learning tool depends on the knowledge of the target problem.

A recently active research field in machine learning is the deep learning framework, which in brief is a generalization of the single layer neural network framework to multiple layers (and hence the nomination deep). (LeCun *et al.*, 2015) One main feature of deep learning is the ability to combine the feature selection and learning steps in a unified framework. However, up to now, there is little theoretical understanding of how it works. How to design an efficient neural network topology for a given problem still remains an art, and a lot of trial and error and ad hoc experience are needed. Moreover, without theoretical understanding, it might be difficult to derive physiological insights from the established neural network, even if it provides a powerful prediction accuracy. Also, while no theoretical guarantee, based on practical experience, it may take a lot of high quality data to make it work properly, which might limit its application to the clinical settings. Despite its theoretical limitations, such deep learning approach has obtained many successes in practical problems. Many variations have been developed in many domains, for example, algorithms to combine audio and video, or the recognition of emotion from multiple modalities,(Mroueh *et al.*, 2015; Kahou *et al.*, 2016) and using auto-encoders limiting the dimension of most inputs,(Said *et al.*, 2017) to name but a few.

While it is possible to learn features from the collected raw data by designing a suitable deep neural network, in the clinical setting it might be beneficial to combine the above-mentioned machine learning techniques in different ways. For example, unsupervised learning techniques could help extract intrinsic genuine features out of the raw data, and allow the deep neural network to focus on the feature organization for the prediction purpose. By doing so, we might render the machine learning framework more interpretable, and might reduce the number of cases needed for the deep learning. Another example of combining supervised and unsupervised learning methods is that in multiple studies of young infants time series (Feldman *et al.*, 2011; Weisman *et al.*, 2011), where it has been shown that neonatal behavior can predict trajectories of neurobehavioral, emotional, and cognitive growth.(Weisman *et al.*, 2011) The effect of early childhood conditions on adult behavior can be translated through time series analysis and machine learning into clear prediction of the adult state, based on the observed neonatal features (Fig. 4). On the other hand, in the clinical setup, we may consider a multi-stage approach (e.g. (Basu Roy *et al.*, 2015)): first, high-risk candidate will be detected using standard tests (e.g. microbiome or FHR monitoring). For high-risk candidate, more complex measures, such as metabolome, will be combined with the microbiome and other methods to produce better classifiers.(Larsen & Dai, 2015) In brief, although it is desirable to have a unified universal framework suitable for analyzing versatile medical data, particularly those for the PS, in practice a careful design of the machine learning framework based on the physiological knowledge and clinical setup is needed.

**OUTLOOK**

As evidenced in the following presentation of the literature, multiple preclinical studies in rodent and fetal sheep models of prenatal stress can now be translated into clinical studies involving pregnant mothers and infants. Methylation and microRNA levels can be correlated with maternal stress, depression and anxiety and with infant's cognitive development to assess the sensitivity of this novel biomarker. Similar data now begin to emerge for histone acetylation in the offspring as a function of maternal stress. We propose that increased stress, depression, and anxiety will have a positive relationship with increased methylation and distinct signatures of histone acetylation. Similarly, there is a growing body of literature on clinical studies validating the advanced FHR techniques for example to detect or predict fetal chronic hypoxia or acidemia at birth. This approach can now be extended to test if these monitoring techniques are useful for detecting PS. We attempted a visualization of the complex multi-scale PS or healthy phenotype in Figure 5.

By integrating multiple non-invasively obtainable sources of information using novel epigenetic, electrophysiological and biochemical approaches via machine learning techniques, the proposed framework could yield progress in maternal–fetal medicine, offering a more precise and truly personalized prediction and new possibilities for designing interventions to improve neurodevelopmental outcomes of pregnancy affected by PS.

**REFERENCES**

Aagaard K, Ma J, Antony KM, Ganu R, Petrosino J & Versalovic J. (2014). The placenta harbors a unique microbiome. *Science translational medicine* **6,** 237ra265.

Adrover E, Berger MA, Perez AA, Tarazi FI & Antonelli MC. (2007). Effects of prenatal stress on dopamine D2 receptor asymmetry in rat brain. *Synapse* **61,** 459-462.

Adrover E, Pallarés ME, Baier CJ, Monteleone MC, Giuliani FA, Waagepetersen HS, Brocco MA, Sonnewald U, A. S & Antonelli MC. (2015). Glutamate neurotransmission is affected in prenatally stressed offspring. *Neurochemistry International* **In press**.

Alonso SJ, Arevalo R, Afonso D & Rodriguez M. (1991). Effects of maternal stress during pregnancy on forced swimming test behavior of the offspring. *Physiol Behav* **50,** 511-517.

Ambeskovic M, Fuchs E, Beaumier P, Gerken M & Metz GA. (2013). Hair trace elementary profiles in aging rodents and primates: links to altered cell homeodynamics and disease. *Biogerontology* **14,** 557-567.

Anegroaie P, Frasch M, Rupprecht S, Antonow-Schlorke I, Muller T, Schubert H, Witte OW & Schwab M. (2016). Development of somatosensory evoked potentials in fetal sheep: Effects of betamethasone. *Acta Physiol (Oxf)*.

Antonow-Schlorke I, Helgert A, Gey C, Coksaygan T, Schubert H, Nathanielsz PW, Witte OW & Schwab M. (2009). Adverse effects of antenatal glucocorticoids on cerebral myelination in sheep. *Obstet Gynecol* **113,** 142-151.

Arnsten AF. (2000). Stress impairs prefrontal cortical function in rats and monkeys: role of dopamine D1 and norepinephrine alpha-1 receptor mechanisms. *Prog Brain Res* **126,** 183-192.

Babenko O, Golubov A, Ilnytskyy Y, Kovalchuk I & Metz GA. (2012). Genomic and epigenomic responses to chronic stress involve miRNA-mediated programming. *PLoS One* **7,** e29441.

Babenko O, Kovalchuk I & Metz GA. (2015). Stress-induced perinatal and transgenerational epigenetic programming of brain development and mental health. *Neurosci Biobehav Rev* **48,** 70-91.

Barros VG, Berger MA, Martijena ID, Sarchi MI, Perez AA, Molina VA, Tarazi FI & Antonelli MC. (2004). Early adoption modifies the effects of prenatal stress on dopamine and glutamate receptors in adult rat brain. *J Neurosci Res* **76,** 488-496.

Barros VG, Duhalde-Vega M, Caltana L, Brusco A & Antonelli MC. (2006a). Astrocyte-neuron vulnerability to prenatal stress in the adult rat brain. *J Neurosci Res* **83,** 787-800.

Barros VG, Rodriguez P, Martijena ID, Perez A, Molina VA & Antonelli MC. (2006b). Prenatal stress and early adoption effects on benzodiazepine receptors and anxiogenic behavior in the adult rat brain. *Synapse* **60,** 609-618.

Basu Roy S, Teredesai A, Zolfaghar K, Liu R, Hazel D, Newman S & Marinez A. (2015). Dynamic hierarchical classification for patient risk-of-readmission. In *Proceedings of the 21th ACM SIGKDD international conference on knowledge discovery and data mining*, pp. 1691-1700. ACM.

Bauer A, Kantelhardt JW, Barthel P, Schneider R, Makikallio T, Ulm K, Hnatkova K, Schomig A, Huikuri H, Bunde A, Malik M & Schmidt G. (2006). Deceleration capacity of heart rate as a predictor of mortality after myocardial infarction: cohort study. *Lancet* **367,** 1674-1681.

Benoit JD, Rakic P & Frick KM. (2015). Prenatal stress induces spatial memory deficits and epigenetic changes in the hippocampus indicative of heterochromatin formation and reduced gene expression. *Behav Brain Res* **281,** 1-8.

Berardino BG, Fesser EA & Canepa ET. (2017). Perinatal protein malnutrition alters expression of miRNA biogenesis genes Xpo5 and Ago2 in mice brain. *Neurosci Lett* **647,** 38-44.

Berger MA, Barros VG, Sarchi MI, Tarazi FI & Antonelli MC. (2002). Long-term effects of prenatal stress on dopamine and glutamate receptors in adult rat brain. *Neurochem Res* **27,** 1525-1533.

Bergman K, Sarkar P, O'Connor TG, Modi N & Glover V. (2007). Maternal stress during pregnancy predicts cognitive ability and fearfulness in infancy. *J Am Acad Child Adolesc Psychiatry* **46,** 1454-1463.

Beydoun H & Saftlas AF. (2008). Physical and mental health outcomes of prenatal maternal stress in human and animal studies: a review of recent evidence. *Paediatr Perinat Epidemiol* **22,** 438-466.

Bingham BC, Sheela Rani CS, Frazer A, Strong R & Morilak DA. (2013). Exogenous prenatal corticosterone exposure mimics the effects of prenatal stress on adult brain stress response systems and fear extinction behavior. *Psychoneuroendocrinology* **38,** 2746-2757.

Black MM, Walker SP, Fernald LCH, Andersen CT, DiGirolamo AM, Lu C, McCoy DC, Fink G, Shawar YR, Shiffman J, Devercelli AE, Wodon QT, Vargas-Baron E, Grantham-McGregor S & Lancet Early Childhood Development Series Steering C. (2017). Early childhood development coming of age: science through the life course. *Lancet* **389,** 77-90.

Blaze J & Roth TL. (2015). Evidence from clinical and animal model studies of the long-term and transgenerational impact of stress on DNA methylation. *Semin Cell Dev Biol* **43,** 76-84.

Bleker LS, Roseboom TJ, Vrijkotte TG, Reynolds RM & de Rooij SR. (2017). Determinants of cortisol during pregnancy - The ABCD cohort. *Psychoneuroendocrinology* **83,** 172-181.

Boersma GJ & Tamashiro KL. (2015). Individual differences in the effects of prenatal stress exposure in rodents. *Neurobiol Stress* **1,** 100-108.

Borovikova LV, Ivanova S, Zhang M, Yang H, Botchkina GI, Watkins LR, Wang H, Abumrad N, Eaton JW & Tracey KJ. (2000). Vagus nerve stimulation attenuates the systemic inflammatory response to endotoxin. *Nature* **405,** 458-462.

Borre YE, O'Keeffe GW, Clarke G, Stanton C, Dinan TG & Cryan JF. (2014). Microbiota and neurodevelopmental windows: implications for brain disorders. *Trends in molecular medicine* **20,** 509-518.

Braithwaite EC, Kundakovic M, Ramchandani PG, Murphy SE & Champagne FA. (2015). Maternal prenatal depressive symptoms predict infant NR3C1 1F and BDNF IV DNA methylation. *Epigenetics : official journal of the DNA Methylation Society* **10,** 408-417.

Braun K, Bock J, Wainstock T, Matas E, Gaisler-Salomon I, Fegert J, Ziegenhain U & Segal M. (2017). Experience-induced transgenerational (re-)programming of neuronal structure and functions: Impact of stress prior and during pregnancy. *Neurosci Biobehav Rev*.

Bravi A, Longtin A & Seely AJ. (2011). Review and classification of variability analysis techniques with clinical applications. *Biomed Eng Online* **10,** 90.

Bygren LO, Kaati G & Edvinsson S. (2001). Longevity determined by paternal ancestors' nutrition during their slow growth period. *Acta Biotheor* **49,** 53-59.

Caldji C, Hellstrom IC, Zhang TY, Diorio J & Meaney MJ. (2011). Environmental regulation of the neural epigenome. *FEBS Lett* **585,** 2049-2058.

Caliskan G & Albrecht A. (2013). Noradrenergic interactions via autonomic nervous system: a promising target for extinction-based exposure therapy? *J Neurophysiol* **110,** 2507-2510.

Cao-Lei L, Dancause KN, Elgbeili G, Laplante DP, Szyf M & King S. (2016a). Pregnant women's cognitive appraisal of a natural disaster affects their children's BMI and central adiposity via DNA methylation: Project Ice Storm. *Early Hum Dev* **103,** 189-192.

Cao-Lei L, Massart R, Suderman MJ, Machnes Z, Elgbeili G, Laplante DP, Szyf M & King S. (2014). DNA methylation signatures triggered by prenatal maternal stress exposure to a natural disaster: Project Ice Storm. *PLoS One* **9,** e107653.

Cao-Lei L, Veru F, Elgbeili G, Szyf M, Laplante DP & King S. (2016b). DNA methylation mediates the effect of exposure to prenatal maternal stress on cytokine production in children at age 13(1/2) years: Project Ice Storm. *Clin Epigenetics* **8,** 54.

Carboni E, Barros VG, Ibba M, Silvagni A, Mura C & Antonelli MC. (2010). Prenatal restraint stress: an in vivo microdialysis study on catecholamine release in the rat prefrontal cortex. *Neuroscience* **168,** 156-166.

Carone BR, Fauquier L, Habib N, Shea JM, Hart CE, Li R, Bock C, Li C, Gu H, Zamore PD, Meissner A, Weng Z, Hofmann HA, Friedman N & Rando OJ. (2010). Paternally induced transgenerational environmental reprogramming of metabolic gene expression in mammals. *Cell* **143,** 1084-1096.

Casati D, Stampalija T, Rizas K, Ferrazzi E, Mastroianni C, Rosti E, Quadrifoglio M & Bauer A. (2014). Assessment of coupling between trans-abdominally acquired fetal ECG and uterine activity by bivariate phase-rectified signal averaging analysis. *PLoS One* **9,** e94557.

Charil A, Laplante DP, Vaillancourt C & King S. (2010). Prenatal stress and brain development. *Brain Res Rev* **65,** 56-79.

Clancy JA, Mary DA, Witte KK, Greenwood JP, Deuchars SA & Deuchars J. (2014). Non-invasive vagus nerve stimulation in healthy humans reduces sympathetic nerve activity. *Brain Stimul* **7,** 871-877.

Cohen S, Kamarck T & Mermelstein R. (1983). A global measure of perceived stress. *Journal of health and social behavior* **24,** 385-396.

Coifman RR, Lafon S, Lee AB, Maggioni M, Nadler B, Warner F & Zucker SW. (2005). Geometric diffusions as a tool for harmonic analysis and structure definition of data: diffusion maps. *Proc Natl Acad Sci U S A* **102,** 7426-7431.

Darnaudery M & Maccari S. (2008). Epigenetic programming of the stress response in male and female rats by prenatal restraint stress. *Brain Res Rev* **57,** 571-585.

Davis EP, Glynn LM, Schetter CD, Hobel C, Chicz-Demet A & Sandman CA. (2007). Prenatal exposure to maternal depression and cortisol influences infant temperament. *J Am Acad Child Adolesc Psychiatry* **46,** 737-746.

Dawes GS, Serra-Serra V, Moulden M & Redman CW. (1994). Dexamethasone and fetal heart rate variation. *Br J Obstet Gynaecol* **101,** 675-679.

Derks JB, Mulder EJ & Visser GH. (1995). The effects of maternal betamethasone administration on the fetus. *Br J Obstet Gynaecol* **102,** 40-46.

Dias BG & Ressler KJ. (2014). Parental olfactory experience influences behavior and neural structure in subsequent generations. *Nat Neurosci* **17,** 89-96.

Diaz R, Fuxe K & Ogren SO. (1997). Prenatal corticosterone treatment induces long-term changes in spontaneous and apomorphine-mediated motor activity in male and female rats. *Neuroscience* **81,** 129-140.

Durosier LD, Green G, Batkin I, Seely AJ, Ross MG, Richardson BS & Frasch MG. (2014). Sampling rate of heart rate variability impacts the ability to detect acidemia in ovine fetuses near-term. *Frontiers in pediatrics* **2,** 38.

Durosier LD, Herry CL, Cortes M, Cao M, Burns P, Desrochers A, Fecteau G, Seely AJ & Frasch MG. (2015). Does heart rate variability reflect the systemic inflammatory response in a fetal sheep model of lipopolysaccharide-induced sepsis? *Physiol Meas* **36,** 2089-2102.

Durosier LD, Xu A, Matushewski B, Cao M, Herry C, Batkin I, Seely A, Ross M, Richardson BS & M.G. F. (2013). Neural signature of cerebral activity of the fetal cholinergic anti-inflammatory pathway derived from heart rate variability *The FASEB journal : official publication of the Federation of American Societies for Experimental Biology* **27**.

Engineer N, Darwin L, Nishigandh D, Ngianga-Bakwin K, Smith SC & Grammatopoulos DK. (2013). Association of glucocorticoid and type 1 corticotropin-releasing hormone receptors gene variants and risk for depression during pregnancy and post-partum. *J Psychiatr Res* **47,** 1166-1173.

Entringer S, Buss C, Kumsta R, Hellhammer DH, Wadhwa PD & Wust S. (2009). Prenatal psychosocial stress exposure is associated with subsequent working memory performance in young women. *Behavioral neuroscience* **123,** 886-893.

Erickson ZT, Falkenberg EA & Metz GA. (2014). Lifespan psychomotor behaviour profiles of multigenerational prenatal stress and artificial food dye effects in rats. *PLoS One* **9,** e92132.

Esler M, Eikelis N, Schlaich M, Lambert G, Alvarenga M, Kaye D, El-Osta A, Guo L, Barton D, Pier C, Brenchley C, Dawood T, Jennings G & Lambert E. (2008). Human sympathetic nerve biology: parallel influences of stress and epigenetics in essential hypertension and panic disorder. *Ann N Y Acad Sci* **1148,** 338-348.

Farrokhi CB, Tovote P, Blanchard RJ, Blanchard DC, Litvin Y & Spiess J. (2007). Cortagine: behavioral and autonomic function of the selective CRF receptor subtype 1 agonist. *CNS Drug Rev* **13,** 423-443.

Fatima M, Srivastav S & Mondal AC. (2017). Prenatal stress and depression associated neuronal development in neonates. *Int J Dev Neurosci* **60,** 1-7.

Feldman R, Magori-Cohen R, Galili G, Singer M & Louzoun Y. (2011). Mother and infant coordinate heart rhythms through episodes of interaction synchrony. *Infant Behav Dev* **34,** 569-577.

Field T, Diego M, Hernandez-Reif M, Vera Y, Gil K, Schanberg S, Kuhn C & Gonzalez-Garcia A. (2004). Prenatal maternal biochemistry predicts neonatal biochemistry. *Int J Neurosci* **114,** 933-945.

Filiberto AC, Maccani MA, Koestler D, Wilhelm-Benartzi C, Avissar-Whiting M, Banister CE, Gagne LA & Marsit CJ. (2011). Birthweight is associated with DNA promoter methylation of the glucocorticoid receptor in human placenta. *Epigenetics : official journal of the DNA Methylation Society* **6,** 566-572.

Fontein-Kuipers YJ, Nieuwenhuijze MJ, Ausems M, Bude L & de Vries R. (2014). Antenatal interventions to reduce maternal distress: a systematic review and meta-analysis of randomised trials. *BJOG* **121,** 389-397.

Francis D, Diorio J, Liu D & Meaney MJ. (1999). Nongenomic transmission across generations of maternal behavior and stress responses in the rat. *Science* **286,** 1155-1158.
22

Frangos E, Ellrich J & Komisaruk BR. (2015). Non-invasive Access to the Vagus Nerve Central Projections via Electrical Stimulation of the External Ear: fMRI Evidence in Humans. *Brain Stimul* **8,** 624-636.

Frasch MG, Muller T, Weiss C, Schwab K, Schubert H & Schwab M. (2009). Heart rate variability analysis allows early asphyxia detection in ovine fetus. *Reproductive sciences (Thousand Oaks, Calif)* **16,** 509-517.

Frasch MG, Muller T, Wicher C, Weiss C, Lohle M, Schwab K, Schubert H, Nathanielsz PW, Witte OW & Schwab M. (2007). Fetal body weight and the development of the control of the cardiovascular system in fetal sheep. *J Physiol* **579,** 893-907.

Frasch MG, Xu Y, Stampalija T, Durosier LD, Herry C, Wang X, Casati D, Seely AJ, Alfirevic Z, Gao X & Ferrazzi E. (2014). Correlating multidimensional fetal heart rate variability analysis with acid-base balance at birth. *Physiol Meas* **35,** L1-12.

Friedman J, Hastie T & Tibshirani R. (2001). *The elements of statistical learning*, vol. 1. Springer series in statistics New York.

Fujioka A, Fujioka T, Ishida Y, Maekawa T & Nakamura S. (2006). Differential effects of prenatal stress on the morphological maturation of hippocampal neurons. *Neuroscience* **141,** 907-915.

Garzoni L, Faure C & Frasch MG. (2013). Fetal cholinergic anti-inflammatory pathway and necrotizing enterocolitis: the brain-gut connection begins in utero. *Frontiers in integrative neuroscience* **7,** 57.

George MS, Ward HE, Jr., Ninan PT, Pollack M, Nahas Z, Anderson B, Kose S, Howland RH, Goodman WK & Ballenger JC. (2008). A pilot study of vagus nerve stimulation (VNS) for treatment-resistant anxiety disorders. *Brain Stimul* **1,** 112-121.

Georgieva A, Papageorghiou AT, Payne SJ, Moulden M & Redman CW. (2014). Phase-rectified signal averaging for intrapartum electronic fetal heart rate monitoring is related to acidaemia at birth. *BJOG* **121,** 889-894.

Gerardin DC, Pereira OC, Kempinas WG, Florio JC, Moreira EG & Bernardi MM. (2005). Sexual behavior, neuroendocrine, and neurochemical aspects in male rats exposed prenatally to stress. *Physiol Behav* **84,** 97-104.

Glover V. (2015). Prenatal stress and its effects on the fetus and the child: possible underlying biological mechanisms. *Adv Neurobiol* **10,** 269-283.

Goldberger AL, Peng CK & Lipsitz LA. (2002). What is physiologic complexity and how does it change with aging and disease? *Neurobiol Aging* **23,** 23-26.

Graatsma EM, Mulder EJ, Vasak B, Lobmaier SM, Pildner von Steinburg S, Schneider KT, Schmidt G & Visser GH. (2012). Average acceleration and deceleration capacity of fetal heart rate in normal pregnancy and in pregnancies complicated by fetal growth restriction. *J Matern Fetal Neonatal Med* **25,** 2517-2522.

Grantham-McGregor S, Cheung YB, Cueto S, Glewwe P, Richter L & Strupp B. (2007). Developmental potential in the first 5 years for children in developing countries. *Lancet* **369,** 60-70.

Grundwald NJ & Brunton PJ. (2015). Prenatal stress programs neuroendocrine stress responses and affective behaviors in second generation rats in a sex-dependent manner. *Psychoneuroendocrinology* **62,** 204-216.

Guyon I & Elisseeff A. (2003). An introduction to variable and feature selection. *Journal of machine learning research* **3,** 1157-1182.

Haberman Y, Tickle TL, Dexheimer PJ, Kim MO, Tang D, Karns R, Baldassano RN, Noe JD, Rosh J, Markowitz J, Heyman MB, Griffiths AM, Crandall WV, Mack DR, Baker SS, Huttenhower C, Keljo DJ, Hyams JS, Kugathasan S, Walters TD, Aronow B, Xavier RJ, Gevers D & Denson LA. (2014). Pediatric Crohn disease patients exhibit specific ileal transcriptome and microbiome signature. *J Clin Invest* **124,** 3617-3633.

Hackman DA, Farah MJ & Meaney MJ. (2010). Socioeconomic status and the brain: mechanistic insights from human and animal research. *Nat Rev Neurosci* **11,** 651-659.

Henry C, Guegant G, Cador M, Arnauld E, Arsaut J, Le Moal M & Demotes-Mainard J. (1995). Prenatal stress in rats facilitates amphetamine-induced sensitization and induces long-lasting changes in dopamine receptors in the nucleus accumbens. *Brain Res* **685,** 179-186.

Herry CL, Cortes M, Wu HT, Durosier LD, Cao M, Burns P, Desrochers A, Fecteau G, Seely AJ & Frasch MG. (2016). Temporal Patterns in Sheep Fetal Heart Rate Variability Correlate to Systemic Cytokine Inflammatory Response: A Methodological Exploration of Monitoring Potential Using Complex Signals Bioinformatics. *PloS one* **11,** e0153515.

Ho DH & Burggren WW. (2010). Epigenetics and transgenerational transfer: a physiological perspective. *J Exp Biol* **213,** 3-16.

Horan DL, Hill LD & Schulkin J. (2000). Childhood sexual abuse and preterm labor in adulthood: an endocrinological hypothesis. *Womens Health Issues* **10,** 27-33.

Hosoi T, Okuma Y & Nomura Y. (2000). Electrical stimulation of afferent vagus nerve induces IL-1beta expression in the brain and activates HPA axis. *Am J Physiol Regul Integr Comp Physiol* **279,** R141-147.

Howerton CL, Morgan CP, Fischer DB & Bale TL. (2013). O-GlcNAc transferase (OGT) as a placental biomarker of maternal stress and reprogramming of CNS gene transcription in development. *Proc Natl Acad Sci U S A* **110,** 5169-5174.

Huhn EA, Lobmaier S, Fischer T, Schneider R, Bauer A, Schneider KT & Schmidt G. (2011). New computerized fetal heart rate analysis for surveillance of intrauterine growth restriction. *Prenat Diagn* **31,** 509-514.

Huizink AC, Mulder EJ & Buitelaar JK. (2004). Prenatal stress and risk for psychopathology: specific effects or induction of general susceptibility? *Psychol Bull* **130,** 115-142.

Iqbal M, Moisiadis VG, Kostaki A & Matthews SG. (2012). Transgenerational effects of prenatal synthetic glucocorticoids on hypothalamic-pituitary-adrenal function. *Endocrinology* **153,** 3295-3307.

Jackson MZ, Gruner KA, Qin C & Tourtellotte WG. (2014). A neuron autonomous role for the familial dysautonomia gene ELP1 in sympathetic and sensory target tissue innervation. *Development* **141,** 2452-2461.

Kahou SE, Bouthillier X, Lamblin P, Gulcehre C, Michalski V, Konda K, Jean S, Froumenty P, Dauphin Y & Boulanger-Lewandowski N. (2016). Emonets: Multimodal deep learning approaches for emotion recognition in video. *Journal on Multimodal User Interfaces* **10,** 99-111.

Kantelhardt JW, Bauer A, Schumann AY, Barthel P, Schneider R, Malik M & Schmidt G. (2007). Phase-rectified signal averaging for the detection of quasi-periodicities and the prediction of cardiovascular risk. *CHAOS* **17,** 015112.

Kapoor A, Leen J & Matthews SG. (2008). Molecular regulation of the hypothalamic-pituitary-adrenal axis in adult male guinea pigs after prenatal stress at different stages of gestation. *J Physiol* **586,** 4317-4326.

Kapoor A & Matthews SG. (2011). Testosterone is involved in mediating the effects of prenatal stress in male guinea pig offspring. *J Physiol* **589,** 755-766.

Katunar MR, Saez T, Brusco A & Antonelli MC. (2010). Ontogenetic expression of dopamine-related transcription factors and tyrosine hydroxylase in prenatally stressed rats. *Neurotox Res* **18,** 69-81.

Kertes DA, Kamin HS, Hughes DA, Rodney NC, Bhatt S & Mulligan CJ. (2016). Prenatal Maternal Stress Predicts Methylation of Genes Regulating the Hypothalamic-Pituitary-Adrenocortical System in Mothers and Newborns in the Democratic Republic of Congo. *Child Dev* **87,** 61-72.

Koenig JI, Elmer GI, Shepard PD, Lee PR, Mayo C, Joy B, Hercher E & Brady DL. (2005). Prenatal exposure to a repeated variable stress paradigm elicits behavioral and neuroendocrinological changes in the adult offspring: potential relevance to schizophrenia. *Behav Brain Res* **156,** 251-261.

Kwan H, Garzoni L, Liu HL, Cao M, Desrochers A, Fecteau G, Burns P & Frasch MG. (2016). VNS in inflammation: systematic review of animal models and clinical studies. *Bioelectronic Medicine* **3,** 1-6.

Laplante DP, Brunet A & King S. (2016). The effects of maternal stress and illness during pregnancy on infant temperament: Project Ice Storm. *Pediatr Res* **79,** 107-113.

Larsen PE & Dai Y. (2015). Metabolome of human gut microbiome is predictive of host dysbiosis. *GigaScience* **4,** 42.

Leclercq S, Mian FM, Stanisz AM, Bindels LB, Cambier E, Ben-Amram H, Koren O, Forsythe P & Bienenstock J. (2017). Low-dose penicillin in early life induces long-term changes in murine gut microbiota, brain cytokines and behavior. *Nat Commun* **8,** 15062.

LeCun Y, Bengio Y & Hinton G. (2015). Deep learning. *Nature* **521,** 436-444.

Lee R, Tamashiro K, Yang X, Purcell R, Harvey A, Willour V, Huo Y, Rongione M, Wand G & Potash J. (2010). Chronic Corticosterone Exposure Increases Expression and Decreases Deoxyribonucleic Acid Methylation of Fkbp5 in Mice. *Endocrinology* **151,** 4332-4343.

Li X, Xu Y, Herry C, Durosier LD, Casati D, Stampalija T, Maisonneuve E, Seely AJ, Audibert F, Alfirevic Z, Ferrazzi E, Wang X & Frasch MG. (2015). Sampling frequency of fetal heart rate impacts the ability to predict pH and BE at birth: a retrospective multi-cohort study. *Physiol Meas* **36,** L1-L12.

Lim JP & Brunet A. (2013). Bridging the transgenerational gap with epigenetic memory. *Trends Genet* **29,** 176-186.

Liu D, Diorio J, Tannenbaum B, Caldji C, Francis D, Freedman A, Sharma S, Pearson D, Plotsky PM & Meaney MJ. (1997). Maternal care, hippocampal glucocorticoid receptors, and hypothalamic-pituitary-adrenal responses to stress. *Science* **277,** 1659-1662.

Liu HL, Butcher J, Romain G, Cao M, Durosier LD, Burns P, Mulon PY, Fecteau G, Desrochers A, Patey N, Garzoni L, Faure C, Stintzi A & Frasch MG. (2015). Fetal gut microbiome diversity is modulated by subclinical ileum inflammation due to systemic endotoxin exposure and by vagal denervation. In *ESPGHAN* pp. 227. J of Pediatric Gastroenterology Nutrition, Amsterdam.

Liu RP, Fang JL, Rong PJ, Zhao Y, Meng H, Ben H, Li L, Huang ZX, Li X, Ma YG & Zhu B. (2013). Effects of electroacupuncture at auricular concha region on the depressive status of unpredictable chronic mild stress rat models. *Evid Based Complement Alternat Med* **2013,** 789674.

Liu S, Gao X, Gao N, Wang X, Fang X, Hu HZ, Wang GD, Xia Y & Wood JD. (2005). Expression of type 1 corticotropin-releasing factor receptor in the guinea pig enteric nervous system. *J Comp Neurol* **481,** 284-298.

Lobmaier S, Ortiz J, Sewald M, Müller A, Schmidt G, Haller B, Oberhoffer R, Schneider K, Giussani D & Wacker-Gussmann A. ((2017)). Influence of gestational diabetes on the fetal autonomic nervous system: A study using phase-rectified signal averaging analysis. . *Ultrasound in Obstet and Gynecol*.

Lobmaier SM, Huhn EA, Pildner von Steinburg S, Muller A, Schuster T, Ortiz JU, Schmidt G & Schneider KT. (2012). Phase-rectified signal averaging as a new method for surveillance of growth restricted fetuses. *J Matern Fetal Neonatal Med* **25,** 2523-2528.

Lobmaier SM, Mensing van Charante N, Ferrazzi E, Giussani DA, Shaw CJ, Muller A, Ortiz JU, Ostermayer E, Haller B, Prefumo F, Frusca T, Hecher K, Arabin B, Thilaganathan B, Papageorghiou AT, Bhide A, Martinelli P, Duvekot JJ, van Eyck J, Visser GH, Schmidt G, Ganzevoort W, Lees CC, Schneider KT & investigators T. (2016). Phase-rectified signal averaging method to predict perinatal outcome in infants with very preterm fetal growth restriction- a secondary analysis of TRUFFLE-trial. *American journal of obstetrics and gynecology* **215,** 630 e631-630 e637.

Lu C, Black MM & Richter LM. (2016). Risk of poor development in young children in low-income and middle-income countries: an estimation and analysis at the global, regional, and country level. *Lancet Glob Health* **4,** e916-e922.

Marschik PB, Pokorny FB, Peharz R, Zhang D, O'Muircheartaigh J, Roeyers H, Bolte S, Spittle AJ, Urlesberger B, Schuller B, Poustka L, Ozonoff S, Pernkopf F, Pock T, Tammimies K, Enzinger C, Krieber M, Tomantschger I, Bartl-Pokorny KD, Sigafoos J, Roche L, Esposito G, Gugatschka M, Nielsen-Saines K, Einspieler C, Kaufmann WE & Group B-PS. (2017). A Novel Way to Measure and Predict Development: A Heuristic Approach to Facilitate the Early Detection of Neurodevelopmental Disorders. *Curr Neurol Neurosci Rep* **17,** 43.

Mastorci F, Vicentini M, Viltart O, Manghi M, Graiani G, Quaini F, Meerlo P, Nalivaiko E, Maccari S & Sgoifo A. (2009). Long-term effects of prenatal stress: changes in adult cardiovascular regulation and sensitivity to stress. *Neuroscience and biobehavioral reviews* **33,** 191-203.

McCallum J, Smith N, Schwab M, Coksaygan T, Reinhardt B, Nathanielsz P & Richardson BS. (2008). Effects of antenatal glucocorticoids on cerebral substrate metabolism in the preterm ovine fetus. *Am J Obstet Gynecol* **198,** 105 e101-109.

McCreary JK, Truica LS, Friesen B, Yao Y, Olson DM, Kovalchuk I, Cross AR & Metz GA. (2016). Altered brain morphology and functional connectivity reflect a vulnerable affective state after cumulative multigenerational stress in rats. *Neuroscience* **330,** 79-89.

Meaney MJ. (2001). Maternal care, gene expression, and the transmission of individual differences in stress reactivity across generations. *Annu Rev Neurosci* **24,** 1161-1192.

Meaney MJ, Aitken DH, van Berkel C, Bhatnagar S & Sapolsky RM. (1988). Effect of neonatal handling on age-related impairments associated with the hippocampus. *Science* **239,** 766-768.

Meltzer-Brody S, Stuebe A, Dole N, Savitz D, Rubinow D & Thorp J. (2011). Elevated corticotropin releasing hormone (CRH) during pregnancy and risk of postpartum depression (PPD). *J Clin Endocrinol Metab* **96,** E40-47.

Moisiadis VG & Matthews SG. (2014a). Glucocorticoids and fetal programming part 1: Outcomes. *Nature reviews Endocrinology* **10,** 391-402.

Moisiadis VG & Matthews SG. (2014b). Glucocorticoids and fetal programming part 2: Mechanisms. *Nature reviews Endocrinology* **10,** 403-411.

Monteleone MC, Adrover E, Pallares ME, Antonelli MC, Frasch AC & Brocco MA. (2014). Prenatal stress changes the glycoprotein GPM6A gene expression and induces epigenetic changes in rat offspring brain. *Epigenetics : official journal of the DNA Methylation Society* **9,** 152-160.

Moog NK, Buss C, Entringer S, Shahbaba B, Gillen DL, Hobel CJ & Wadhwa PD. (2016). Maternal Exposure to Childhood Trauma Is Associated During Pregnancy With Placental-Fetal Stress Physiology. *Biol Psychiatry* **79,** 831-839.

Mroueh Y, Marcheret E & Goel V. (2015). Deep multimodal learning for audio-visual speech recognition. In *Acoustics, Speech and Signal Processing (ICASSP), 2015 IEEE International Conference on*, pp. 2130-2134. IEEE.

Mueller BR & Bale TL. (2008). Sex-specific programming of offspring emotionality after stress early in pregnancy. *J Neurosci* **28,** 9055-9065.

Mulder EJ, Derks JB & Visser GH. (1997). Antenatal corticosteroid therapy and fetal behaviour: a randomised study of the effects of betamethasone and dexamethasone. *Br J Obstet Gynaecol* **104,** 1239-1247.

Mulder EJ, Robles de Medina PG, Huizink AC, Van den Bergh BR, Buitelaar JK & Visser GH. (2002). Prenatal maternal stress: effects on pregnancy and the (unborn) child. *Early Hum Dev* **70,** 3-14.

Mulligan CJ, D'Errico NC, Stees J & Hughes DA. (2012). Methylation changes at NR3C1 in newborns associate with maternal prenatal stress exposure and newborn birth weight. *Epigenetics : official journal of the DNA Methylation Society* **7,** 853-857.

Murgatroyd C, Patchev AV, Wu Y, Micale V, Bockmuhl Y, Fischer D, Holsboer F, Wotjak CT, Almeida OF & Spengler D. (2009). Dynamic DNA methylation programs persistent adverse effects of early-life stress. *Nat Neurosci* **12,** 1559-1566.

Negron-Oyarzo I, Neira D, Espinosa N, Fuentealba P & Aboitiz F. (2015). Prenatal Stress Produces Persistence of Remote Memory and Disrupts Functional Connectivity in the Hippocampal-Prefrontal Cortex Axis. *Cereb Cortex* **25,** 3132-3143.

Nolvi S, Karlsson L, Bridgett DJ, Korja R, Huizink AC, Kataja EL & Karlsson H. (2016). Maternal prenatal stress and infant emotional reactivity six months postpartum. *J Affect Disord* **199,** 163-170.

O'Connor TG, Caprariello P, Blackmore ER, Gregory AM, Glover V, Fleming P & Team AS. (2007). Prenatal mood disturbance predicts sleep problems in infancy and toddlerhood. *Early human development* **83,** 451-458.

O'Connor TG, Heron J, Golding J & Glover V. (2003). Maternal antenatal anxiety and behavioural/emotional problems in children: a test of a programming hypothesis. *Journal of child psychology and psychiatry, and allied disciplines* **44,** 1025-1036.

O'Keane V, Dinan TG, Scott L & Corcoran C. (2005). Changes in hypothalamic-pituitary-adrenal axis measures after vagus nerve stimulation therapy in chronic depression. *Biol Psychiatry* **58,** 963-968.

Pallares ME, Adrover E, Baier CJ, Bourguignon NS, Monteleone MC, Brocco MA, Gonzalez-Calvar SI & Antonelli MC. (2013a). Prenatal maternal restraint stress exposure alters the reproductive hormone profile and testis development of the rat male offspring. *Stress* **16,** 429-440.

Pallares ME, Baier CJ, Adrover E, Monteleone MC, Brocco MA & Antonelli MC. (2013b). Age-dependent effects of prenatal stress on the corticolimbic dopaminergic system development in the rat male offspring. *Neurochem Res* **38,** 2323-2335.

Pastor V, Antonelli MC & Pallares ME. (2017). Unravelling the Link Between Prenatal Stress, Dopamine and Substance Use Disorder. *Neurotox Res* **31,** 169-186.

Pavlov VA & Tracey KJ. (2012). The vagus nerve and the inflammatory reflex--linking immunity and metabolism. *Nature reviews Endocrinology* **8,** 743-754.

Pavlov VA & Tracey KJ. (2015). Neural circuitry and immunity. *Immunologic research* **63,** 38-57.

Pavlov VA & Tracey KJ. (2017). Neural regulation of immunity: molecular mechanisms and clinical translation. *Nat Neurosci* **20,** 156-166.

Pena DF, Childs JE, Willett S, Vital A, McIntyre CK & Kroener S. (2014). Vagus nerve stimulation enhances extinction of conditioned fear and modulates plasticity in the pathway from the ventromedial prefrontal cortex to the amygdala. *Front Behav Neurosci* **8,** 327.

Porges SW. (1995). Cardiac vagal tone: a physiological index of stress. *Neurosci Biobehav Rev* **19,** 225-233.

Porges SW. (2009). The polyvagal theory: new insights into adaptive reactions of the autonomic nervous system. *Cleve Clin J Med* **76 Suppl 2,** S86-90.

Provencal N, Suderman MJ, Guillemin C, Massart R, Ruggiero A, Wang D, Bennett AJ, Pierre PJ, Friedman DP, Cote SM, Hallett M, Tremblay RE, Suomi SJ & Szyf M. (2012). The signature of maternal rearing in the methylome in rhesus macaque prefrontal cortex and T cells. *The Journal of neuroscience : the official journal of the Society for Neuroscience* **32,** 15626-15642.

Pusalkar M, Suri D, Kelkar A, Bhattacharya A, Galande S & Vaidya VA. (2016). Early stress evokes dysregulation of histone modifiers in the medial prefrontal cortex across the life span. *Dev Psychobiol* **58,** 198-210.

Rakers F, Frauendorf V, Rupprecht S, Schiffner R, Bischoff SJ, Kiehntopf M, Reinhold P, Witte OW, Schubert H & Schwab M. (2013). Effects of early- and late-gestational maternal stress and synthetic glucocorticoid on development of the fetal hypothalamus-pituitary-adrenal axis in sheep. *Stress* **16,** 122-129.

Rakers F, Rupprecht S, Dreiling M, Bergmeier C, Witte OW & Schwab M. (2017). Transfer of maternal psychosocial stress to the fetus. *Neurosci Biobehav Rev*.

Rieger M, Pirke KM, Buske-Kirschbaum A, Wurmser H, Papousek M & Hellhammer DH. (2004). Influence of stress during pregnancy on HPA activity and neonatal behavior. *Ann N Y Acad Sci* **1032,** 228-230.

Rivolta MW, Stampalija T, Casati D, Richardson BS, Ross MG, Frasch MG, Bauer A, Ferrazzi E & Sassi R. (2014). Acceleration and deceleration capacity of fetal heart rate in an in-vivo sheep model. *PloS one* **9,** e104193.

Rodriguez N, Mayer N & Gauna HF. (2007). Effects of prenatal stress on male offspring sexual maturity. *Biocell* **31,** 67-74.

Roseboom TJ, van der Meulen JH, Ravelli AC, Osmond C, Barker DJ & Bleker OP. (2001). Effects of prenatal exposure to the Dutch famine on adult disease in later life: an overview. *Mol Cell Endocrinol* **185,** 93-98.

Roweis ST & Saul LK. (2000). Nonlinear dimensionality reduction by locally linear embedding. *Science* **290,** 2323-2326.

Said AB, Mohamed A, Elfouly T, Harras K & Wang ZJ. (2017). Multimodal deep learning approach for joint EEG-EMG data compression and classification. In *Wireless Communications and Networking Conference (WCNC), 2017 IEEE*, pp. 1-6. IEEE.

Santavirta T, Santavirta N & Gilman SE. (2017). Association of the World War II Finnish Evacuation of Children With Psychiatric Hospitalization in the Next Generation. *JAMA Psychiatry*.

Schwab M, Coksaygan T, Rakers F & Nathanielsz PW. (2012). Glucocorticoid exposure of sheep at 0.7 to 0.75 gestation augments late-gestation fetal stress responses. *Am J Obstet Gynecol* **206,** 253 e216-222.

Schwab M, Schmidt K, Roedel M, Mueller T, Schubert H, Anwar MA & Nathaniels PW. (2001). Non-linear changes of electrocortical activity after antenatal betamethasone treatment in fetal sheep. *J Physiol* **531,** 535-543.

Seely AJE & Newman KD. (2016). *Monitoring Variability and Complexity at the Bedside*. Springer International Publishing.

Senat MV, Minoui S, Multon O, Fernandez H, Frydman R & Ville Y. (1998). Effect of dexamethasone and betamethasone on fetal heart rate variability in preterm labour: a randomised study. *Br J Obstet Gynaecol* **105,** 749-755.

Shono T & Suita S. (2003). Disturbed pituitary-testicular axis inhibits testicular descent in the prenatal rat. *BJU Int* **92,** 641-643.

Silvagni A, Barros VG, Mura C, Antonelli MC & Carboni E. (2008). Prenatal restraint stress differentially modifies basal and stimulated dopamine and noradrenaline release in the nucleus accumbens shell: an 'in vivo' microdialysis study in adolescent and young adult rats. *Eur J Neurosci* **28,** 744-758.

Silveira PP & Manfro GG. (2015). Retrospective studies. *Adv Neurobiol* **10,** 251-267.

Stampalija T, Casati D, Monasta L, Sassi R, Rivolta MW, Muggiasca ML, Bauer A & Ferrazzi E. (2016). Brain sparing effect in growth-restricted fetuses is associated with decreased cardiac acceleration and deceleration capacities: a case-control study. *BJOG* **123,** 1947-1954.

Stampalija T, Casati D, Montico M, Sassi R, Rivolta MW, Maggi V, Bauer A & Ferrazzi E. (2015). Parameters influence on acceleration and deceleration capacity based on trans-abdominal ECG in early fetal growth restriction at different gestational age epochs. *Eur J Obstet Gynecol Reprod Biol* **188,** 104-112.

Stampalija T, Signaroldi M, Mastroianni C, Rosti E, Signorelli V, Casati D & Ferrazzi EM. (2012). Fetal and maternal heart rate confusion during intra-partum monitoring: comparison of trans-abdominal fetal electrocardiogram and Doppler telemetry. *J Matern Fetal Neonatal Med* **25,** 1517-1520.

Talmon & Wu. ((2017)). Latent common manifold learning with alternating diffusion: analysis and applications. *Applied and computational harmonic analysis*.

Teh AL, Pan H, Chen L, Ong ML, Dogra S, Wong J, MacIsaac JL, Mah SM, McEwen LM, Saw SM, Godfrey KM, Chong YS, Kwek K, Kwoh CK, Soh SE, Chong MF, Barton S, Karnani N, Cheong CY, Buschdorf JP, Stunkel W, Kobor MS, Meaney MJ, Gluckman PD & Holbrook JD. (2014). The effect of genotype and in utero environment on interindividual variation in neonate DNA methylomes. *Genome research* **24,** 1064-1074.

Tenenbaum JB, de Silva V & Langford JC. (2000). A global geometric framework for nonlinear dimensionality reduction. *Science* **290,** 2319-2323.

Van den Berg GJ & Pinger PR. (2014). A Validation Study of Transgenerational Effects of Childhood Conditions on the Third Generation Offspring's Economic and Health Outcomes Potentially Driven by Epigenetic Imprinting. *The Institute for the Study of Labor (IZA)* **IZA Discussion Paper No. 7999**.

Van den Bergh BR, Mulder EJ, Mennes M & Glover V. (2005). Antenatal maternal anxiety and stress and the neurobehavioural development of the fetus and child: links and possible mechanisms. A review. *Neurosci Biobehav Rev* **29,** 237-258.

Van den Hove DL, Steinbusch HW, Scheepens A, Van de Berg WD, Kooiman LA, Boosten BJ, Prickaerts J & Blanco CE. (2006). Prenatal stress and neonatal rat brain development. *Neuroscience* **137,** 145-155.

Van Leeuwen P, Geue D, Thiel M, Cysarz D, Lange S, Romano MC, Wessel N, Kurths J & Gronemeyer DH. (2009). Influence of paced maternal breathing on fetal-maternal heart rate coordination. *Proceedings of the National Academy of Sciences of the United States of America* **106,** 13661-13666.

van Os J & Selten JP. (1998). Prenatal exposure to maternal stress and subsequent schizophrenia. The May 1940 invasion of The Netherlands. *The British journal of psychiatry : the journal of mental science* **172,** 324-326.

Vangeel EB, Izzi B, Hompes T, Vansteelandt K, Lambrechts D, Freson K & Claes S. (2015). DNA methylation in imprinted genes IGF2 and GNASXL is associated with prenatal maternal stress. *Genes Brain Behav* **14,** 573-582.

Vezina-Audette R, Herry C, Burns P, Frasch M, Chave E & Theoret C. (2016). Heart rate variability in relation to stress in the Asian elephant (Elephas maximus). *Can Vet J* **57,** 289-292.

Vidal AC, Benjamin Neelon SE, Liu Y, Tuli AM, Fuemmeler BF, Hoyo C, Murtha AP, Huang Z, Schildkraut J, Overcash F, Kurtzberg J, Jirtle RL, Iversen ES & Murphy SK. (2014). Maternal stress, preterm birth, and DNA methylation at imprint regulatory sequences in humans. *Genet Epigenet* **6,** 37-44.

Wadhwa PD, Entringer S, Buss C & Lu MC. (2011). The contribution of maternal stress to preterm birth: issues and considerations. *Clin Perinatol* **38,** 351-384.

Wadhwa PD, Sandman CA, Porto M, Dunkel-Schetter C & Garite TJ. (1993). The association between prenatal stress and infant birth weight and gestational age at birth: a prospective investigation. *Am J Obstet Gynecol* **169,** 858-865.

Walker SP, Wachs TD, Gardner JM, Lozoff B, Wasserman GA, Pollitt E & Carter JA. (2007). Child development: risk factors for adverse outcomes in developing countries. *Lancet* **369,** 145-157.

Wallen A, Zetterstrom RH, Solomin L, Arvidsson M, Olson L & Perlmann T. (1999). Fate of mesencephalic AHD2-expressing dopamine progenitor cells in NURR1 mutant mice. *Exp Cell Res* **253,** 737-746.

Ward IL & Weisz J. (1984). Differential effects of maternal stress on circulating levels of corticosterone, progesterone, and testosterone in male and female rat fetuses and their mothers. *Endocrinology* **114,** 1635-1644.

Weaver IC, Cervoni N, Champagne FA, D'Alessio AC, Sharma S, Seckl JR, Dymov S, Szyf M & Meaney MJ. (2004). Epigenetic programming by maternal behavior. *Nat Neurosci* **7,** 847-854.

Weaver IC, Champagne FA, Brown SE, Dymov S, Sharma S, Meaney MJ & Szyf M. (2005). Reversal of maternal programming of stress responses in adult offspring through methyl supplementation: altering epigenetic marking later in life. *The Journal of neuroscience : the official journal of the Society for Neuroscience* **25,** 11045-11054.

Weinstock M. (1997). Does prenatal stress impair coping and regulation of hypothalamic-pituitary-adrenal axis? *Neurosci Biobehav Rev* **21,** 1-10.

Weinstock M. (2001). Alterations induced by gestational stress in brain morphology and behaviour of the offspring. *Prog Neurobiol* **65,** 427-451.

Weinstock M. (2008). The long-term behavioural consequences of prenatal stress. *Neuroscience and biobehavioral reviews* **32,** 1073-1086.

Weinstock M. (2017). Prenatal stressors in rodents: Effects on behavior. *Neurobiol Stress* **6,** 3-13.

Weisman O, Magori-Cohen R, Louzoun Y, Eidelman AI & Feldman R. (2011). Sleep-wake transitions in premature neonates predict early development. *Pediatrics* **128,** 706-714.

Werner EA, Myers MM, Fifer WP, Cheng B, Fang Y, Allen R & Monk C. (2007). Prenatal predictors of infant temperament. *Dev Psychobiol* **49,** 474-484.

Yaka R, Salomon S, Matzner H & Weinstock M. (2007). Effect of varied gestational stress on acquisition of spatial memory, hippocampal LTP and synaptic proteins in juvenile male rats. *Behav Brain Res* **179,** 126-132.

Yang J, Han H, Cao J, Li L & Xu L. (2006). Prenatal stress modifies hippocampal synaptic plasticity and spatial learning in young rat offspring. *Hippocampus* **16,** 431-436.

Yao Y, Robinson AM, Zucchi FC, Robbins JC, Babenko O, Kovalchuk O, Kovalchuk I, Olson DM & Metz GA. (2014). Ancestral exposure to stress epigenetically programs preterm birth risk and adverse maternal and newborn outcomes. *BMC Med* **12,** 121.

Yasuda K, Matsunaga T, Moritani T, Nishikino M, Gu N, Yoshinaga M, Nagasumi K, Yamamura T, Aoki N & Tsuda K. (2004). T393C polymorphism of GNAS1 associated with the autonomic nervous system in young, healthy Japanese subjects. *Clin Exp Pharmacol Physiol* **31,** 597-601.

Ylikoski J, Lehtimaki J, Pirvola U, Makitie A, Aarnisalo A, Hyvarinen P & Ylikoski M. (2017). Non-invasive vagus nerve stimulation reduces sympathetic preponderance in patients with tinnitus. *Acta Otolaryngol***,** 1-9.

Zaidan H & Gaisler-Salomon I. (2015). Prereproductive stress in adolescent female rats affects behavior and corticosterone levels in second-generation offspring. *Psychoneuroendocrinology* **58,** 120-129.

Zaidan H, Leshem M & Gaisler-Salomon I. (2013). Prereproductive stress to female rats alters corticotropin releasing factor type 1 expression in ova and behavior and brain corticotropin releasing factor type 1 expression in offspring. *Biol Psychiatry* **74,** 680-687.

Zucchi FC, Yao Y, Ilnytskyy Y, Robbins JC, Soltanpour N, Kovalchuk I, Kovalchuk O & Metz GA. (2014). Lifetime stress cumulatively programs brain transcriptome and impedes stroke recovery: benefit of sensory stimulation. *PLoS One* **9,** e92130.

Zuena AR, Mairesse J, Casolini P, Cinque C, Alema GS, Morley-Fletcher S, Chiodi V, Spagnoli LG, Gradini R, Catalani A, Nicoletti F & Maccari S. (2008). Prenatal restraint stress generates two distinct behavioral and neurochemical profiles in male and female rats. *PLoS ONE* **3,** e2170.

**Figure legends**

**Figure 1.** A schematic representation of the dopaminergic pathways in the rat brain. Dopaminergic neurons can be divided into four groups: nigrostratial, mesolimbic, mesocortical, and tuberohypophyseal systems (see text for details). b Schematic representation of the alterations in the dopaminergic system in the adult rat brain of prenatally restrained stressed rat males. Note that impairments of the dopaminergic system are observed mainly in limbic areas of prenatally stressed rats. From Baier et al, 2012.

**Figure 2.** Temporal profile of a fetal HRV composite measure (principal component analysis, PCA) tracks accurately the pro-inflammatory cytokine IL-6 (pink bars). The red line represents the PCA for IL-6 deviations from baseline in response to intravenous LPS-injection at 0 h (LPS, n = 10 fetal lambs); LPS is the immune stimulus from gram negative bacteria ("infection"). The green line represents the PCA for IL-6 deviations from baseline for control (saline-injected) animals (CON, n = 7). Lightly shaded areas correspond to the confidence intervals around the mean. The baseline value is represented by the dotted blue line.

**Figure 3.** Analytical flow for the derivation of a heart rate variability time series.

**Figure 4.** Proposed framework for future studies of PS effects in a hybrid animal/human experimental design to accelerate biomarker discovery and validation. Note the multi-scale approach spanning several species and data acquisition techniques with varying spatio-temporal resolution. This requires machine learning techniques to derive at risk assessment algorithms for detection of PS exposure and prediction of neurodevelopmental outcomes.

**Figure 5.** A complex multi-scale phenotype of the healthy or prenatally stressed individual. Can the rapidly advancing machine learning techniques help distinguish such individual phenotypes based on all its features across the scales of observations, from microbiome, over to epigenetic landscape to the heart rate (HR) time series?

**Table 1. Description of heart rate variability domains.***

| Domain | Features |
|---|---|
| **Statistical** | The statistical domain consists of statistical measures (mean, standard deviation, Gaussian, and so on) describing the data distribution. It assumes the data originates from a stochastic process. |
| **Geometric** | The geometric domain describes the properties related to the shape of the dataset in space. This includes, in a deterministic system, grid counting, heart rate turbulence, spatial filling index, and Poincaré and recurrence plots. |
| **Energetic** | The energetic domain describes the features related to the energy or the power of the data, such as frequency, periodicity, and irreversibility in time. |
| **Informational** | The informational domain describes the degree of complexity and irregularity in the elements of a time series, such as distance from periodicity or from a reference model. It includes various measures of entropy (compression, fuzzy, multiscale, and so on). |
| **Invariant** | The invariant domain describes the properties of a system that demonstrate fractality or other attributes that do not change over either space or time. Included are scaling exponents, fluctuation analysis, and multifractal exponents. |

* Domains suggested for continuous individualized multiorgan variability analysis (CIMVA platform). Reproduced with permission from Durosier *et al.*, 2015.

**Table 2. Measures included in each domain of fetal heart rate variability for continuous individualized multiorgan variability analysis.**

| Domain | Fetal heart rate variability measure |
|---|---|
| **Statistical** | Coefficient of variation (based on intervals) |
| | Form factor |
| | Interquartile range |
| | Kurtosis |
| | Lee parameter |
| | Mean value |
| | Mean rate |
| | Mean of the differences |
| | Root mean square of successive differences of R-R intervals |
| | Skewness |
| | Standard deviation |
| | Standard deviation of the differences |
| | Symbolic dynamics: modified conditional entropy, non-uniform case |
| | Symbolic dynamics: modified conditional entropy, uniform case |
| | Symbolic dynamics: forbidden words, non-uniform case |
| | Symbolic dynamics: forbidden words, uniform case |
| | Symbolic dynamics: Shannon entropy, non-uniform case |
| | Symbolic dynamics: Shannon entropy, uniform case |
| | Symbolic dynamics: percentage of 0 variations sequences, non-uniform case |
| | Symbolic dynamics: percentage of 0 variations sequences, uniform case |
| | Symbolic dynamics: percentage of 1 variations sequences, non-uniform case |
| | Symbolic dynamics: percentage of 1 variations sequences, uniform case |
| | Symbolic dynamics: percentage of 2 variations sequences, non-uniform case |
| | Symbolic dynamics: percentage of 2 variations sequences, uniform case |
| **Geometric** | Dynamic moment of the second order |
| | Dynamical moment of the third order along the principal bisector |
| | Dynamical moment of the third order along the secondary bisector |
| | Dynamical moment of the third order along the x-axis |
| | Dynamical moment of the third order along the y-axis |
| | Finite growth rates |
| | Grid transformation feature: grid count |
| | Poincaré plot SD1 |
| | Poincaré plot SD2 |
| | Poincaré plot cardiac sympathetic index |
| | Poincaré plot cardiac vagal index |
| | Recurrence quantification analysis: average diagonal line |
| | Recurrence quantification analysis: maximum diagonal line |
| | Recurrence quantification analysis: maximum vertical line |
| | Recurrence quantification analysis: determinism/recurrences |
| | Recurrence quantification analysis: percentage of determinism |
| | Recurrence quantification analysis: percentage of laminarity |
| | Recurrence quantification analysis: percentage of recurrences |
| | Recurrence quantification analysis: Shannon entropy of the diagonals |
| | Recurrence quantification analysis: Shannon entropy of the vertical lines |
| | Recurrence quantification analysis: trapping time |
| **Energetic** | Low frequency/high frequency ratio |
| | Low frequency (LF) power |
| | High frequency (HF) power |
| | Hjorth parameters: activity |
| | Hjorth parameters: complexity |
| | Hjorth parameters: mobility |
| | Multifractal spectrum cumulant of the first order |
|  | Multifractal spectrum cumulant of the second order |
|  | Multifractal spectrum cumulant of the third order |
|  | Multiscale time irreversibility asymmetry index |
|  | Plotkin and Swamy energy operator: average energy |
|  | Teager energy operator: average energy |
|  | Very low frequency power |
|  | Wavelet area under the curve |
| **Informational** | Allan factor distance from a Poisson distribution |
|  | Fano factor distance from a Poisson distribution |
|  | Fuzzy entropy |
|  | Grid transformation feature: AND similarity index |
|  | Grid transformation feature: time delay similarity index |
|  | Grid transformation feature: weighted similarity index |
|  | Index of variability distance from a Poisson distribution |
|  | Kullback-Leibler permutation entropy |
|  | Multiscale entropy |
|  | Predictive feature: error from an autoregressive model |
|  | Sample entropy |
|  | Shannon entropy |
|  | Similarity index of the distributions |
| **Invariant** | Correlation dimension global exponent |
|  | Detrended fluctuation analysis: a1 |
|  | Detrended fluctuation analysis: a2 |
|  | Detrended fluctuation analysis: area under the curve |
|  | Detrended fluctuation analysis: overall a |
|  | Diffusion entropy |
|  | Embedding scaling exponent |
|  | Kolmogorov-Sinai entropy |
|  | Higuchi scaling exponent |
|  | Largest Lyapunov exponent |
|  | Power Law (based on frequency) slope x2 |
|  | Power Law (based on frequency) y-intercept x2 |
|  | Power Law (based on frequency) x-intercept x2 |
|  | Power Law (based on frequency) goodness of fit x2 |
|  | Power Law (based on histogram) slope |
|  | Power Law (based on histogram) y-intercept |
|  | Power Law (based on histogram) x-intercept |
|  | Power Law (based on histogram) goodness of fit |
|  | Rescaled detrended range analysis |
|  | Scale-dependent Lyapunov exponent slope |
|  | Scale-dependent Lyapunov exponent mean value |
|  | Scaled windowed variance |

*, LF =[0.04-0.2 Hz]; **, HF=[0.2-2 Hz]; ***, VLF=[0.001-0.04 Hz].

#, CIMVA= continuous individualized multiorgan variability analysis.

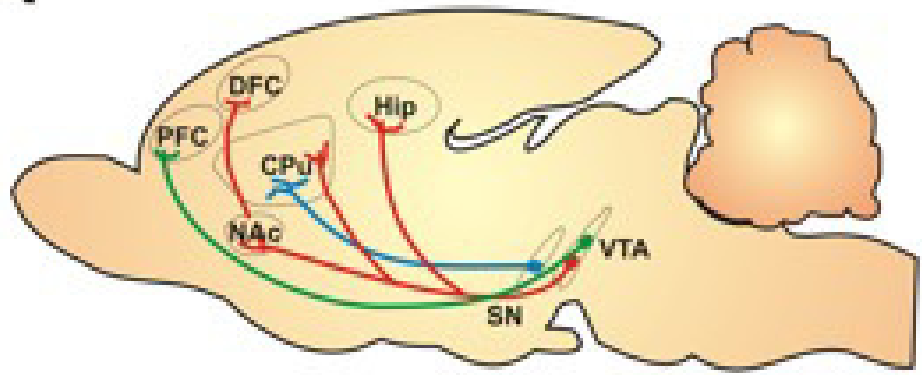
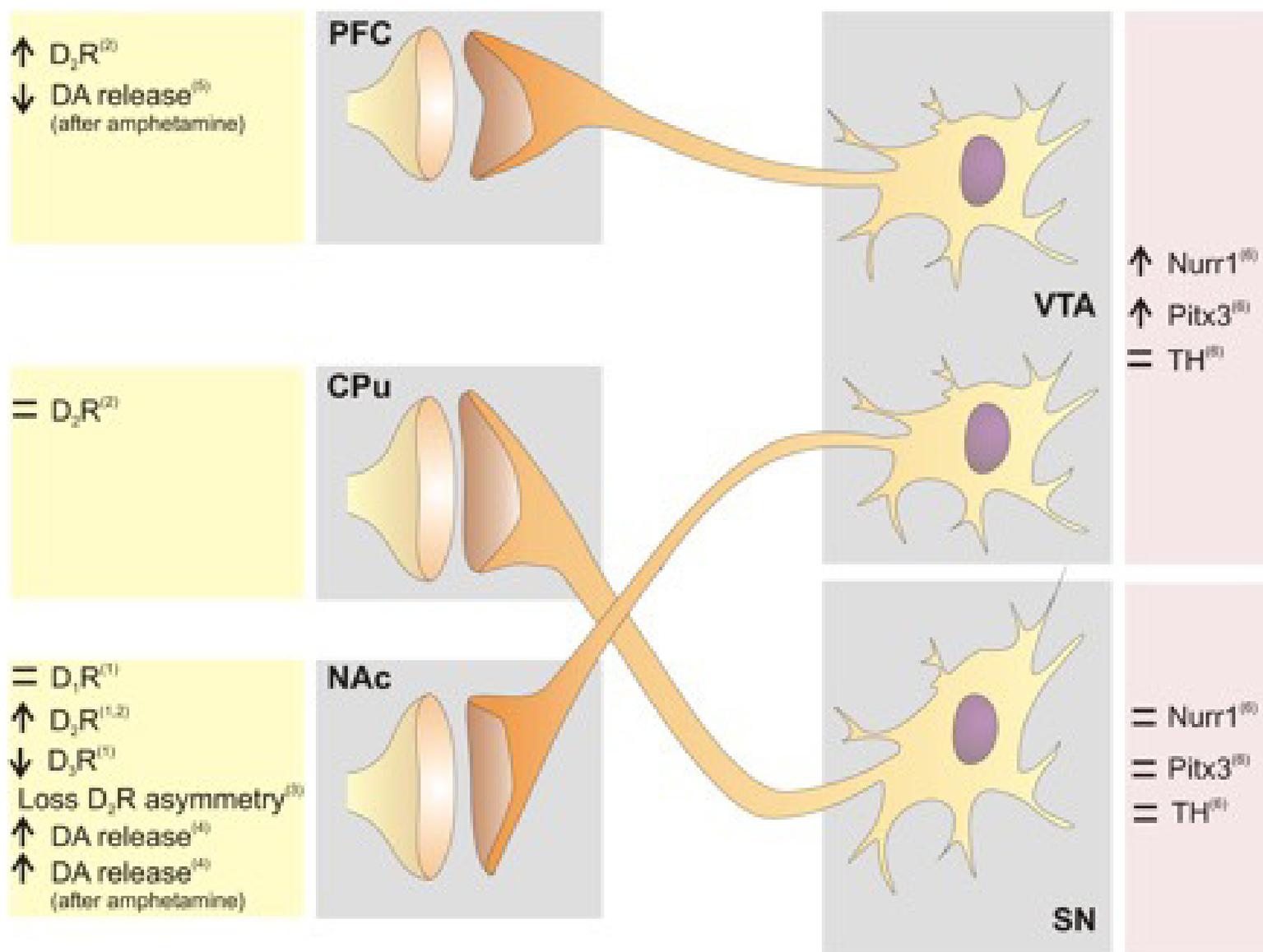

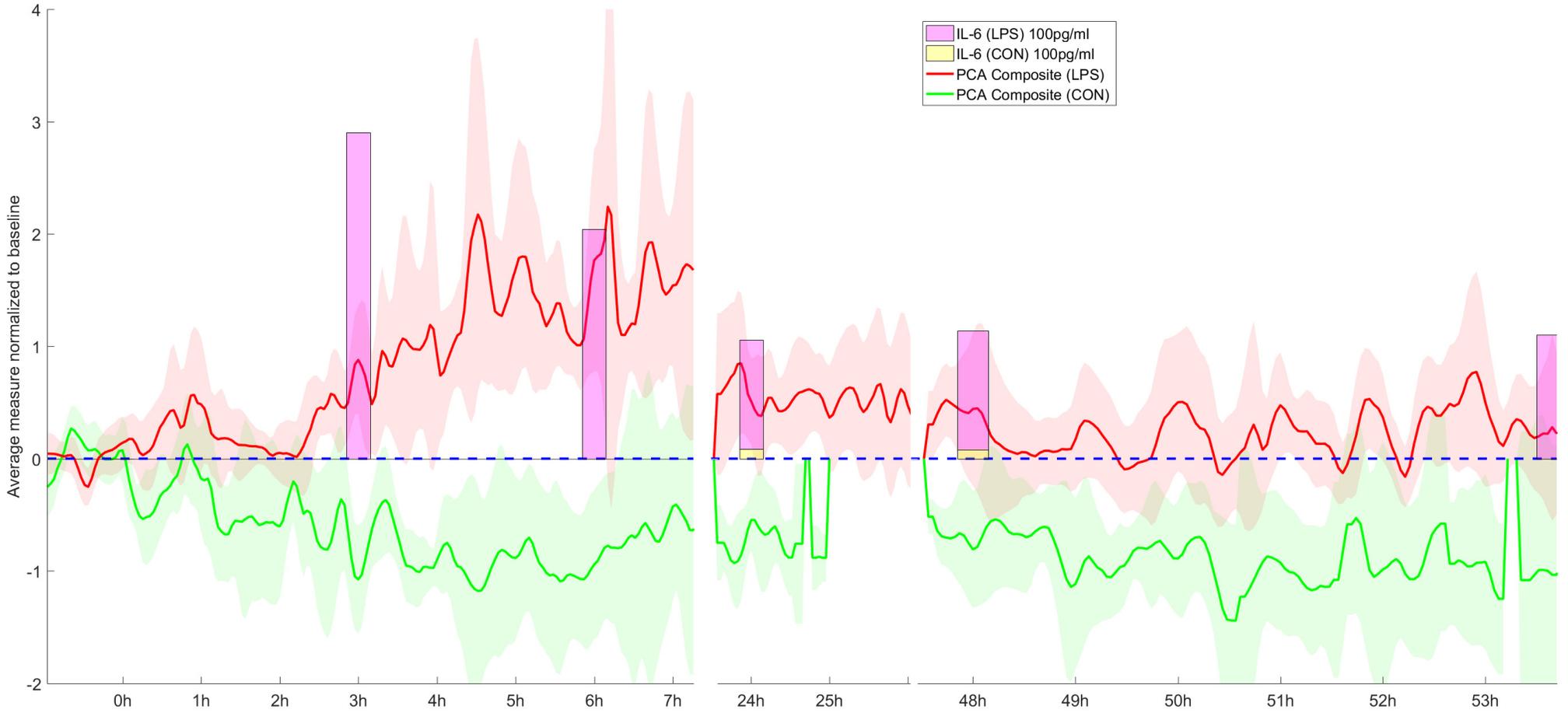

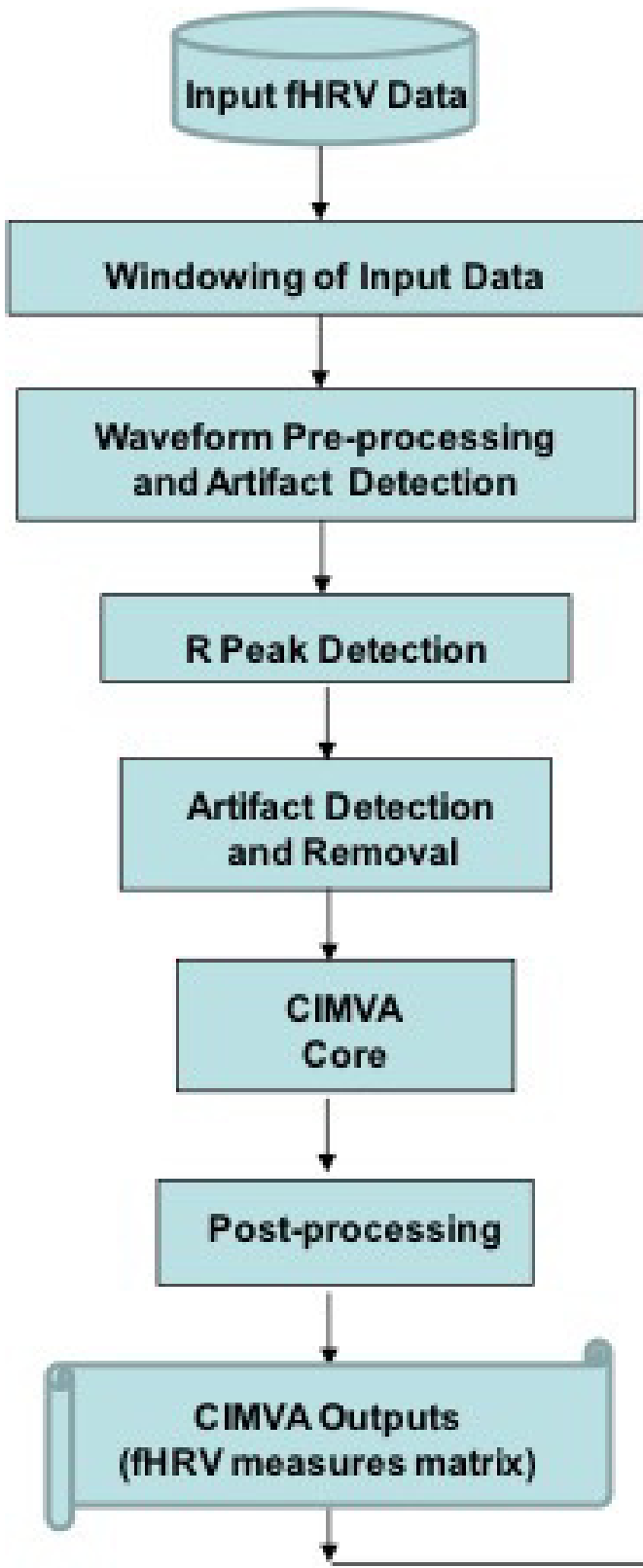

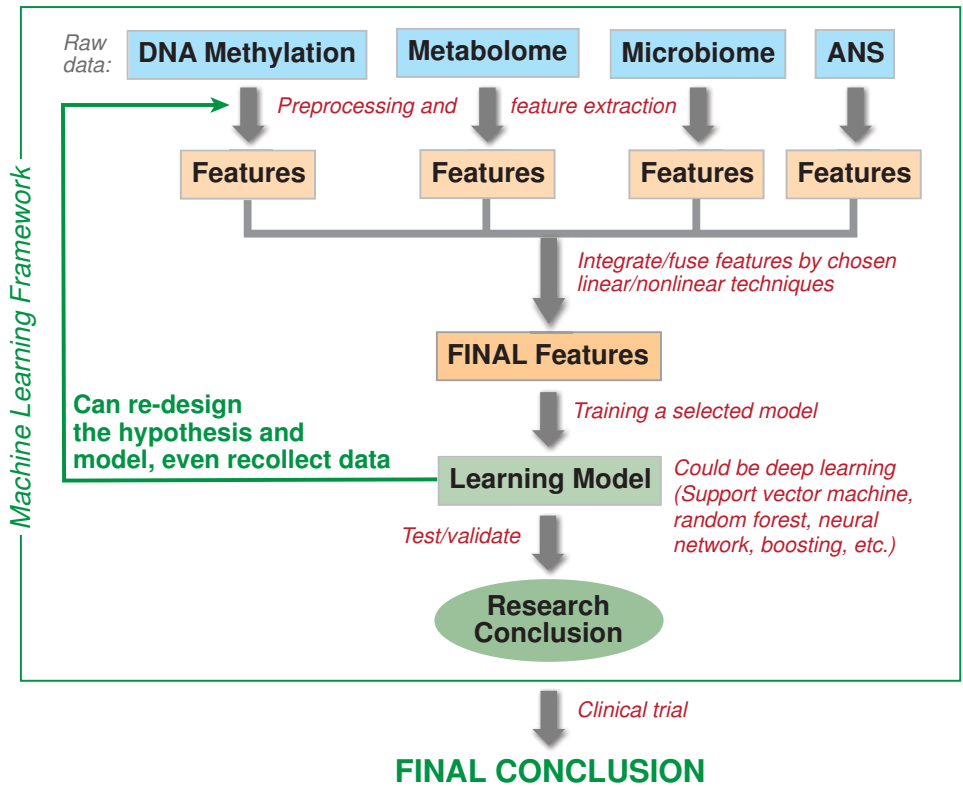

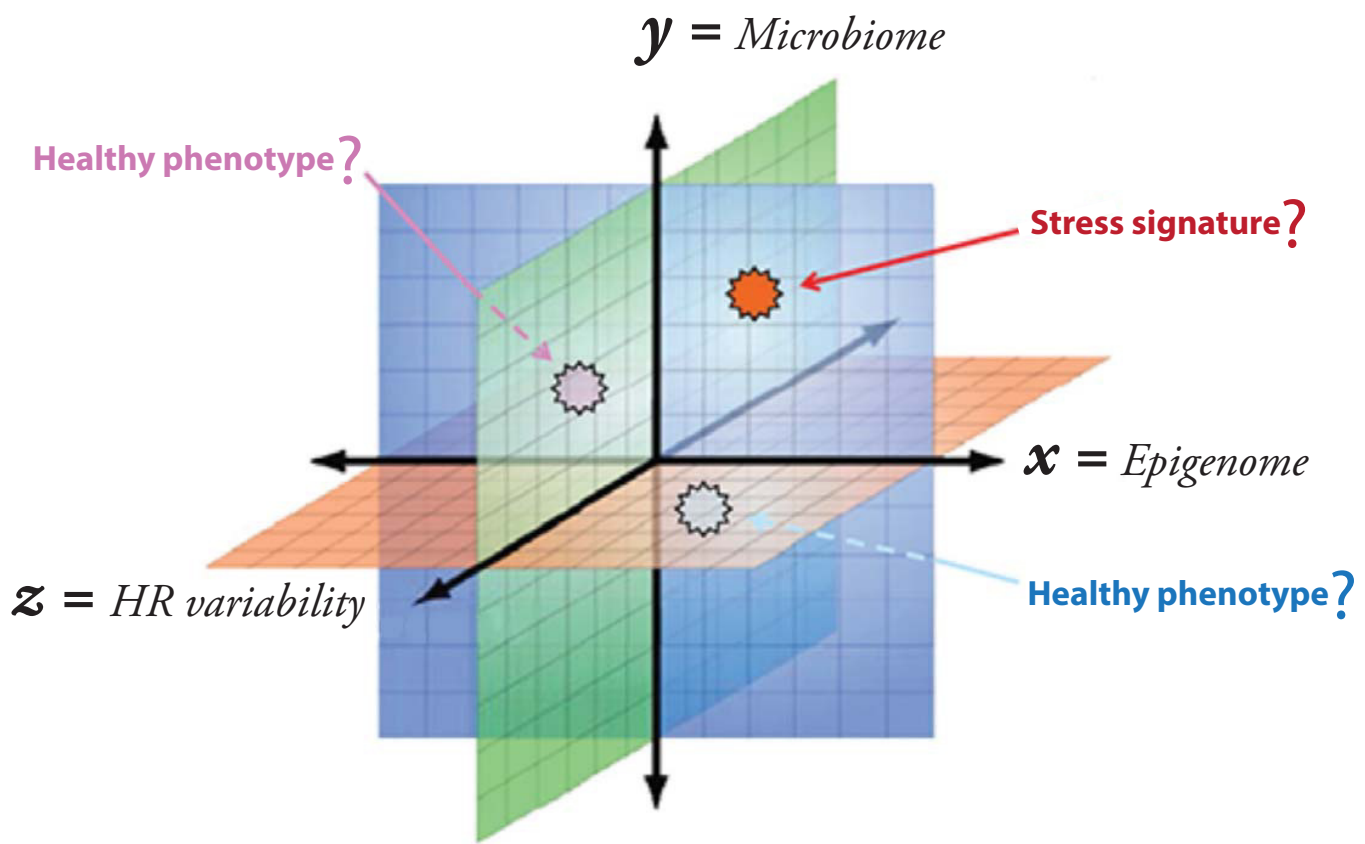